\documentclass[preprint,prb]{revtex4}

\pdfoutput=1
\usepackage{graphicx}
\newcommand{\ev}{\mathbf{E}}
\newcommand{\av}{\mathbf{A}}
\newcommand{\prob}{Problem} 

\usepackage{epsfig}
\usepackage{amsmath}
\usepackage{boxedminipage}
\usepackage{enumerate}

\begin{document}

\title{Student understanding of symmetry and Gauss's law of electricity}
\author{Chandralekha Singh}
\affiliation{Department of Physics and Astronomy, University
of Pittsburgh, Pittsburgh, Pennsylvania 15260}

\begin{abstract}
We investigate the difficulties that students in calculus-based introductory
physics courses have with the concepts of symmetry, electric field, and
electric flux which are important for applying Gauss's law. 
The determination of the electric field using Gauss's law requires
determining the symmetry of a particular charge distribution and
predicting the direction of the electric field everywhere if a high symmetry
exists. Effective application of Gauss's law implicitly requires 
understanding the principle of superposition for electric fields. Helping
students learn when Gauss's law can be readily applied to determine the
strength of the electric field, and then helping them learn to determine
the appropriate shape of Gaussian surfaces if sufficient symmetry exists,
can help develop their reasoning and problem-solving skills. We administered
free-response and multiple-choice questions and conducted interviews with
individual students using a think-aloud protocol to elucidate the
difficulties that students have with the concepts of symmetry, electric field,
and electric flux. We also developed a
multiple-choice test that targets these conceptual issues to obtain quantitative
information about their difficulties and administered it to 541 students in
the introductory calculus-based physics courses and to upper-level
undergraduates in an electricity and magnetism course and to
graduate students enrolled in a TA seminar course. We find that
undergraduate students have many common difficulties with these concepts. 
\end{abstract}

\maketitle

\section{Introduction}

A major goal of most calculus-based introductory physics courses is to help
students develop problem solving and reasoning skills.\cite{alan,meltzer,redish,zollman} 
Gauss's law of electricity is an important topic in the second semester of most calculus-based introductory
physics courses. Learning to reason whether Gauss's law can be exploited in a
particular situation to determine the electric field, without having to
evaluate complicated integrals, can provide an excellent context for helping
students develop a good grasp of symmetry considerations. Unfortunately,
students often memorize a collection of formulas for the magnitude of the
electric field for various geometries, without
paying attention to symmetry considerations. Students apply these formulas
without being able to differentiate between electric field and electric
flux. They have difficulty identifying situations where Gauss's law is
useful and overgeneralize results obtained for a highly symmetric charge
distributions to situations where they are not applicable. Most textbooks
do not sufficiently emphasize symmetry considerations or the chain of
reasoning required to determine if Gauss's law is useful for calculating
the electric field.

To investigate student understanding of the concepts of symmetry, electric
field, and electric flux, we administered free-response and multiple-choice
questions and conducted interviews with 15 individual students using a
think-aloud protocol.\cite{chi} We then developed a multiple-choice test with 25 conceptual questions that addresses these issues and administered it to 541
students in the introductory calculus-based physics courses in eight
different classes to obtain a quantitative understanding of the
nature of the difficulties. The test was also administered as a pre- and
posttest to
undergraduates enrolled in an upper-level electricity and magnetism (E\&M)
course and to graduate students enrolled in a TA seminar course. The tests
and interviews explored the extent to which students have become proficient
in exploiting symmetry and in making conceptual predictions about the
magnitude and direction of the electric field for a given charge
distribution using Coulomb's or Gauss's laws. The test also explores
whether students can distinguish between electric field and electric flux,
identify situations in which Gauss's law can readily be used to calculate
the electric field strength from the information about the electric flux,
and the shapes of the Gaussian surfaces that would be appropriate in those
cases.

\section{Previous Investigations related to Electricity and Magnetism}

Investigation of student difficulties related to a particular physics
concept is important for designing instructional strategies to reduce
them.\cite{maloney,lillian,beichner,rainson,belcher} Prior investigations
related to electricity and magnetism have included difficulties with general
introductory concepts, electrical circuits, and superposition of
the electric field.\cite{maloney,lillian,beichner,rainson,belcher} Maloney
et al.\cite{maloney} developed and administered a 32 question multiple-choice
test (the Conceptual Survey of Electricity and Magnetism) that
surveys many important concepts covered in the introductory physics courses
and is suitable for both calculus- and algebra-based courses. They found that
students have common difficulties with fundamental concepts related to
electricity and magnetism. McDermott et al.\cite{lillian} performed an
in-depth investigation of the difficulties students have with electrical
circuits and developed exemplary tutorials and inquiry-based curriculum
to significantly reduce these difficulties among introductory physics
students and pre- and in-service teachers.\cite{lillian2} Beichner et
al.\cite{beichner} have developed conceptual assessments related to
electrical circuits. Viennot et al.\cite{rainson} investigated difficulties
with the superposition of electric fields by administering written
questions. Belcher et al.\cite{belcher} and Belloni and
Christian\cite{mario} have developed visualization tools to improve student
understanding of physics concepts including those related to electricity and
magnetism.

\section{Method}

Our investigation of student difficulties in discerning symmetry and
applying Gauss's law was performed using two methods: the design and
administration of free-response and multiple-choice questions to elicit
difficulties in a particular context and in-depth audio-taped
interviews with individual students using a think-aloud protocol\cite{chi}
while they solved those problems. A principal advantage of written tests is
that they can be administered to large student populations. Multiple-choice tests are easy and economical to
administer and to grade, have objective scoring, and are amenable to
statistical analysis that can be used to compare student populations or
instructional methods. The main drawback is that thought processes are not
revealed completely by the answers alone. However, when combined with 
interviews with a subset of individual students, well-designed tests can serve as
excellent educational tools. Conceptual multiple-choice
tests have already been designed to assess student understanding of force,
energy and momentum, and electricity and
magnetism.\cite{maloney,beichner,mdt,fmce,singh} They show that students'
knowledge of physics is often fragmented and context-dependent and that
students share common difficulties.

\section{Background}

Gauss's law allows us to relate the net electric flux through a closed
surface to the net charge enclosed by the surface. 
\begin{equation}
\Phi_E=Q_{\rm enclosed}/\epsilon_0. \label{gauss}
\end{equation}
Equation~(\ref{gauss}) implies that if we know the net electric flux through a
closed surface, we can readily find the net charge inside it. And if we
know the net charge inside a closed surface, we can readily find the net
electric flux through it.

In general, Eq.~(\ref{gauss}) does not mean that we can use Gauss's
law to readily find the magnitude of the electric field $\vert \ev
\vert$ at a point. Only in situations where the charge distribution has
very high symmetry can we find $\vert \ev \vert$ from the net electric
flux
$\Phi_E$. Although there are only three types of symmetry (spherical,
cylindrical, and planar) for which Gauss's law can readily be exploited to
determine the electric field at various points from the information about
the electric flux, students need help in identifying when these symmetries
are present. 

The net electric flux through a closed surface is given by
\begin{equation}
\Phi_E
=\!\oint\! \ev \cdot d\av =\!\oint\! \vert \ev \vert \cos \theta 
|d\av|.\label{relation}
\end{equation}
The net electric flux $\Phi_E$ over a closed (Gaussian) surface can be
exploited to determine the electric field magnitude $\vert \ev \vert$ at an
arbitrary point $P$ on the surface only if the following conditions
are satisfied:

\begin{enumerate}

\item We can determine the exact direction of $\ev$ relative to the area
vector at every point on the closed surface by symmetry (only $\theta=0$,
$180^\circ$ or $\pm 90^\circ$ are associated with sufficiently high symmetry).

\item 
In some cases, we can divide the closed surface into subsections (for each
subsection the electric flux can be readily calculated for example, the side
and two caps of a cylinder) such that one of the following is
true:
\begin{enumerate}
\item $\vert \ev \vert$ is the same everywhere on the subsection due to the
symmetry of the charge distribution.

\item $\ev$ and the area vector (outward normal to the surface) are
perpendicular ($\theta=90^\circ$) so that there is no electric flux through
that subsection.
\end{enumerate}
\end{enumerate}
Thus, to determine if the information about the net electric flux through a
closed surface can be exploited to determine 
$\vert \ev \vert$ at a point P, we may choose a Gaussian
(closed imaginary) surface such that it contains the point $P$ where we want to determine $\vert\ev \vert$ and
$\vert \ev \vert \cos \theta$ is known (by symmetry) to have a
constant value on each subsection of the surface so that it can be
pulled out of the flux integral in Eq.~(\ref{relation}). Then $\int\!|
d\av \vert$ equals the total area of the subsection of the surface.

Although students can calculate the electric field without regard to
symmetry using Coulomb's law and the principle of superposition, 
determining the electric field using Gauss's law requires an explicit focus on
the symmetry of the charge distribution. The principle of superposition is
also a prerequisite for employing Gauss's law successfully to determine the
electric field (albeit implicit in the actual application of Gauss's law)
and helps determine if sufficient charge symmetry exists in a particular
situation. In addition to considerations of symmetry, the area vector
and the electric flux are new concepts that are introduced in the context
of Gauss's law. Students must be able to distinguish the electric flux from
the the electric field, a task that is very difficult. Students
need to learn that the electric flux is a scalar which is positive
and negative depending on the relative directions of the electric field
and area vector, while the electric field at a point is a vector. 
They need to learn that the electric flux and the electric field have different
dimensions, that Gauss's law applies only to closed surfaces, and that for
any closed surface information about the net enclosed charge is sufficient
to determine the net electric flux through it. On the other hand, the
determination of the electric field at various points due to a charge
distribution depends on the way in which the charges are 
distributed and can depend on both the charges inside and outside the
closed surface. For example, the electric field at various points on a
Gaussian surface may vary from point to point, even though the net electric
flux is zero (and hence the net enclosed charge is zero).

\section{Discussion of Student Difficulties}

We first administered free-response and multiple-choice questions and
interviewed individual students in several calculus-based introductory
physics courses about concepts related to symmetry and Gauss's law.
These investigations provided useful insight into common difficulties
students have with these concepts and also provided guidelines for the
development of a 25 question multiple-choice test that was then used as a tool
to obtain quantitative information about the extent to which student have
common difficulties with these concepts. Appendix~\ref{app:summary} provides a summary of
the development of the multiple-choice test. The final version of the test
is in Appendix~\ref{app:test}. This 50 minute test was administered to 541
students in the calculus-based introductory physics courses in eight
different classes at the University of Pittsburgh. Five additional
student volunteers were interviewed (in addition to those interviewed
earlier for a total of 15 individual interviews) to obtain a greater
understanding of their difficulties. Two of these classes were honors courses. The average score
on the test was
49\%. Table~\ref{tab1} shows the percent of students who selected the choices
(a)--(e) on \prob s 1--25. The correct responses are
italicized. Although some questions have a strong
single distractor, others have several distractor choices that are equally
popular. In eight of the test questions, students had to
determine the correctness of three statements. Students identified the
correctness of some of the statements but not others as shown in Table~\ref{tab1} (we
will discuss these issues in detail later in this section). 
The reliability coefficient\cite{nitko}
$\alpha$, which is a measure of the internal consistency of the
test, is $0.8$, which is considered good by the standards of test design.
The point biserial discrimination~\cite{nitko} (PBD) quantifies the ability of a question to
discriminate between students who did well overall and those who
did not. This discrimination index for 16 questions was more
than $0.4$. Only 4 questions had a PBD value less than $0.3$ (only one
question with less than 0.2), which is also good by the standards of test
design.\cite{nitko} The reliability coefficient and discrimination indices
show that the conceptual difficulties with symmetry and Gauss's law found
using the test are meaningful.

Table~\ref{tab2} shows the concepts that were covered and the questions in the multiple-choice test that addressed them. 
The table provides only one of the several ways questions can be classified. Some of the categories in Table~\ref{tab2} are 
subsets of other categories. We have found these subcategories convenient
for classifying student difficulties. The
categorization of concepts in Table~\ref{tab2} is based on student
difficulties and does not necessarily reflect the way experts would
categorize those problems. For example, one of the concept categories in
which we placed \prob s 13 and 15 is ``distinguishing between the
electric field and electric flux.'' Although the word ``flux" is never
explicitly mentioned in these questions, our investigation shows that
students often believed that because there was no charge enclosed inside the
Gaussian surfaces that are exclusively contained in the hollow region, the
electric field must be zero everywhere inside the hollow region. They were
often interpreting zero net electric flux through a closed surface to imply
zero electric field at all points of the surface. We classified
\prob s 19, 21, and 23 in the same concept category ``distinguishing
between the electric field and electric flux.'' As shown in Table~\ref{tab1}, the
most common misconception in \prob~19 was due to confusion between the
electric field and the electric flux. In \prob~21 the most common
incorrect response was choice (a) because students believed that in order to
determine the net electric flux through a closed surface, knowledge of the
charge enclosed was not sufficient and charges had to be symmetrically
distributed (a requirement for determining the electric field). The most
common incorrect response for \prob~23 was choice (a), which also arises
due to the difficulty in distinguishing between the electric field and flux.
Similarly, we placed \prob~14 in the category ``recognizing
symmetry to determine if it is easy to exploit Gauss's law \ldots," because
many students incorrectly believed that even in experiment 1, Gauss's law
can be used to infer that the magnitude of the electric field is the same at
points A, B, and C. 

\subsection{The electric charge and electric flux are
scalars}

The most common difficulty with \prob~1 was mistakenly thinking that the
electric flux and/or electric charge are vectors. In interviews,
students justified their response about why the electric flux is a vector by
using the following facts: The expression for flux involves a scalar product
of two vectors. Instead of identifying
$\cos \theta$ as the angle between the electric field and the area vector,
many students concluded that the flux is a vector because it involves a
$\cos \theta$. Students pointed to the fact that the electric flux can
have both positive and negative signs. When asked if it would make
sense to say that the electric flux points at 
$30^\circ$ south of west, students often avoided a direct response. Their
response implied that for a
physical quantity to be a vector, it was not necessary to be able to specify the exact direction. Rather,
because the electric field lines ``going out" of
a closed surface contribute positively and those ``going in" contribute negatively to the total electric flux through a closed surface, flux must be a vector. 
To justify why the electric charge
should be a vector, students often made similar claims that positive charges point outward and negative charges point inward. It was
clear from the responses that students were often referring to the electric field but calling it
``charge.''

\subsection{The principle of superposition}

The performance of many students was closely tied with their understanding of the principle of superposition.
For example, \prob s~2--4 are related to symmetry and require the use
of the superposition principle to compare the electric field at various
points for a given charge distribution. Many students had difficulty with
the principle of superposition and could not differentiate between the
electric field due to individual charges at a point and the net electric
field. Interviews suggest that some believed that only the nearest charge
will contribute to the electric field at a point. Others believed that the
magnitude of the electric field at the desired points in \prob~2
(and 3) should be the same because they were the same
perpendicular distance from the straight line joining the three charges.
Some students provided more detailed reasoning. Instead of viewing it as a
problem involving the addition of three electric field vectors, these
students often made guesses by looking at the distances of points A, B, and
C from the three charges and hoping that the electric field will somehow
work out to be the same at the three points. They claimed that in
\prob~2, point A is closer to one charge and farther away from the other
two charges than point B which is equidistant from the two charges and not
as far away from the third charge as point A. Therefore, the electric field
at points A and B will be the same if we take into account all the three
charges. Because charge is uniformly distributed on the finite sheet in
\prob~3, this type of confusion was even more prevalent. The most common
distractor in \prob~4 implies a similar difficulty. Many students
believed that if charges are uniformly distributed on an insulating
equilateral triangle, the magnitude of the electric field will be the same
everywhere on a concentric imaginary triangle.

\subsection{The electric field inside a hollow
non-conducting object}

\prob s 13, 15, and 25 assess student understanding of the electric
field inside hollow non-conducting objects of different shapes due to
charges on their surface or charges outside. \prob~13 was the most
difficult question on the test and only
21\% of the introductory students responded correctly. 
55\% of the students believed that the electric field inside a non-conducting
hollow cube with charge uniformly distributed on its surface will be zero
everywhere. Interviews suggest that some students believed that the hollow
region inside is always shielded from the charges on the surface or charges
outside. This notion of shielding was retained by the students despite being
reminded by the interviewer that the object on which charges are distributed
is not conducting. Some students explicitly said that the net effect of all
the charges outside must work out to be zero everywhere inside the hollow
region. One student went on to claim that he has always been amazed at
how Gauss's law can be used to prove that the electric field in the
hollow region inside a closed object is always zero everywhere, a result
that appears to be counterintuitive to him. Some students even drew
spherical or cubic Gaussian surfaces inside a hollow cube to argue that
because there is no charge enclosed, the electric field will be zero
everywhere according to Gauss's law. Similar to \prob~13, the most
common difficulty with \prob~15 was assuming that the electric field
inside the sphere in experiment 1 is also zero everywhere. In interviews
and free-response questions, students used reasoning similar to \prob~13.

For \prob~25, choices (b)--(d) were popular due to the
difficulties with the principle of superposition and the electric field
inside a hollow non-conducting sphere. Interviews suggest that students who
believed that the electric field at point A is not zero often thought that
the sphere with the uniform surface charge $+Q$ will produce a larger
electric field at that point because one of its sides is only a distance $L$
away compared to the point charge $+Q$ which is a distance $2L$ away. At the
end of the interview, in response to the query by some students, the
interviewer discussed with them that for point A, the charge on the sphere
can be thought of as a point charge at the center of the sphere. Students
often noted that this fact is very non-intuitive because of the proximity of
one end of the sphere to point A. Students who claimed that the net electric
field at point B is zero often refered to the shielding of the inside of the
sphere from the charges on the sphere and charges outside of the sphere
(similar to \prob s~13 and 15) when they were explicitly asked by the
interviewer why the point charge near the sphere does not produce an
electric field at point B. Even when the interviewer reminded students that
the sphere was non-conducting, they often maintained that the point charge
cannot have any influence inside the sphere. Some students said that they
could not explain exactly why the non-conducting sphere will produce
shielding, but that they remember that the electric field must somehow cancel
in the hollow region for all shapes and charge distributions. Further
prodding showed that due to a lack of thorough understanding, these students
were often overgeneralizing or confusing two different facts: the
symmetry argument that shows (using Gauss's law) that the electric field for
a sphere with a uniform surface charge is zero everywhere inside regardless
of whether the sphere is conducting or insulating, and/or the fact that
the electric field inside a conductor is zero in equilibrium regardless of
the shape of the conductor. 

\subsection{The underlying symmetry of a charge
distribution}

Many students have difficulty realizing that it is the symmetry of the
charge distribution (and not the symmetry of the object on which the charges
are embedded) that is important in determining whether Gauss's law can be
applied to calculate the electric field at a point. For example, students
had to determine when a Gaussian surface would be convenient for determining
the electric field at a point on its surface in \prob s~6, 11, and 22. In
\prob s~11 and 22 they had to identify the shape of appropriate Gaussian
surfaces that would make it easy to use Gauss's law to calculate the
electric field due to an infinite uniform sheet of charge and line of charge
respectively. In \prob~11 the most common distractor was (e), and many students
believed that all surfaces will work because they are all symmetric.
However, the calculation for the sphere is not easy because the area vector
and the electric field make different angles for different infinitesimal
areas on the sphere. In \prob~6 students were asked to evaluate the
validity of three general statements without being given a specific
charge distribution that produced the electric field. Many students chose (e) and believed that the Gaussian surface must be chosen to take
advantage of the symmetry of the object enclosed, regardless of how the
charges are distributed on that object. 

The idea of whether the symmetry of the charge distribution or the symmetry
of the object on which charges are distributed is important for being able to
determine the electric field using Gauss's law is also explored in \prob s~10 and 14--16. In \prob~10 Gauss's law can readily be used to
determine the electric field in cases (i) and (iii) because of the spherical
symmetry of the charge distribution on nonpolarizable objects but not in
case (ii). In \prob s~14--16 students were
presented with an insulating sphere on which there are six point
charges distributed in a way that the adjacent charges are equidistant.
Although the charges are on a spherical object, the charge distribution does
not have spherical symmetry. Interviews and written responses suggest that
many students incorrectly believed that a spherical symmetry exists in this
case for exploiting Gauss's law to calculate the electric field readily. The
most common difficulty with \prob~14 was assuming that the magnitude of
the electric field in experiment 1 would be the same at the three
points shown. In interviews and free-response questions, some students
explained this respnse by claiming that because the six point charges and the
three points are symmetrically situated, the field magnitude must be the same at the three points; others explained
their response by claiming that for a point outside, the six point charges
on the sphere can be thought to be point charges at the center of the sphere. \prob~16 is one of the most difficult questions
on the test. It was easy for most students to rule out (i), but it
was difficult for them to evaluate the validity of (ii) and (iii)
because of the difficulty of recognizing the underlying symmetry of the
charge distribution. 

\prob s~17 and 22 also probe the extent to which students can discern the
underlying symmetry of the charge distribution. In \prob~17 the last
three distractors were popular and students often believed that we can
use Gauss's law to find the electric field at a point outside due
to a cube or finite cylinder with uniform surface charge. In the interviews
students sometime recalled using Gauss's law for these surfaces. More
prodding showed that they were either confusing the fact that those surfaces
can be used as Gaussian surfaces for appropriate charge distributions or the
fact that for an infinite cylinder (but not a finite cylinder) it is
possible to exploit Gauss's law to find the electric field. It
appears that many students have not thought carefully about the principle of
superposition and its implication for the electric field due to a charge
distribution and were applying memorized knowledge whose applicability
was forgotten. In \prob~22
students had to choose the Gaussian surfaces that would help them determine
the electric field at point P readily due to the infinite line of charge.
All of the alternative choices were selected with an almost uniform
frequency. Students were often unsure about the symmetry concepts relevant
for making appropriate decisions and those who chose option (c) were often
quite confident that the magnitude of the electric field due to the infinite
line must be the same at every point on the cube as well.

\subsection{Electric field and electric flux}

\prob s~7, 8, and 9 are related to electric flux and the distinction
between the electric field and flux. The most common distractor in \prob~8 was (d). Some students were quite assertive in their interview and
incorrectly claimed that if the magnitude of the electric flux through one
closed surface is smaller than another, the magnitude of the electric field
at points on the first surface must be smaller too. All the distractors were
popular for \prob~9. Interviews suggest that students who chose
(c) often believed that although the net charge enclosed is the same for
surfaces A and B, the electric flux through surface A must be larger because
it is closer to the positive charge at the center. Interviews also suggest
that students who selected option (d) often believed that although the net
charge enclosed is the same for surfaces A and B, the electric flux through
surface B must be larger because it has a larger area.

Similar difficulty in differentiating and relating the electric field and
electric flux was manifested in response to \prob s~12, 18, 21, 22, and 23.
The most common difficulty with \prob~12 was assuming that both electric
flux and electric field can be determined using Gauss's law; 30\% of the
students chose option (c). Interviews suggest that
students were often confident about their choice of (c)
because the surface area of the Gaussian sphere was given and they believed
that
$\Phi=\vert \ev \vert A$ is always true (instead of $\Phi =\oint \ev
\cdot d\av $). \prob~18 is more abstract than most other questions
in which a specific physical situation is explicitly given. In this
question students had to evaluate the validity of three statements and the
main difficulty was not being clear about the relation and distinction
between the electric field and electric flux. The most common difficulty
with \prob~19 was assuming that the electric field is zero at point B
on the side surface of the cube, although the problem statement explicitly
mentions that the cube is in a uniform electric field of 20\,N/C. In
interviews and free-response questions, some students explicitly claimed
that the area vector of the side surface is perpendicular to the direction
of the electric field lines. Therefore, the electric field must be zero at
point B. This kind of confusion between the electric field at a point and
the contribution to the electric flux from a certain area was found in other
questions as well.

Response to \prob s~21 and 22 suggest that many students are not
comfortable with the statement of Gauss's law that relates the net flux
through a closed surface to the net charge enclosed. They have difficulty
differentiating between the electric flux through a closed surface and the
electric field at a point on the surface. For example, in \prob~21
many students chose (i) (or (i) and (iii)) and claimed in the interviews
and free-response questions that only those surfaces can be used to
determine the net electric flux through them because the other surfaces did
not have the correct symmetry. The most common distractor in \prob~23
was (a), which was chosen by 33\% of the students. These students believed
that the net charge enclosed in a region is largest if the number of field
lines penetrating the region is greatest. They did not pay attention to the
direction of the electric field lines which is crucial for determining the
net flux through a closed surface and hence the net charge enclosed using
Gauss's law. 

\subsection{Other Difficulties}

Students were sometimes unsure about the distinction between open and closed surfaces and that Gauss's law is only applicable to 
closed surfaces. 
\prob s~5, 21, and 22 at least partly assess whether students
understand this distinction. Some students incorrectly believed that Gauss's
law applies to any symmetrical surface even if it is not closed. For
example, in response to \prob~21, these students claimed that the
electric flux $\Phi$ due to an infinitely long line of charge (with uniform
linear charge density $\lambda$) is 
$\lambda L/\epsilon_0$ even for the two-dimensional square sheet. 

In response to \prob~24 many students selected some of the statements as 
correct but only 29\% identified that all three statements are correct. 
Many had difficulty determining the validity of statement (ii) which can be checked by drawing a Gaussian surface in region A that
includes the center and using Gauss's law.
In the interview a student insisted that there can be a
point charge in region A even if the electric field is zero. When asked to explain, the student drew a positive point charge with the electric
field lines radially outward and said: ``well\,\ldots\ the field due to a
point charge will cancel out because it points in all directions.'' It was
clear from the explanation that the student was confused about the electric
field line representation.

\subsection{Performance of Upper-Level Undergraduates}

We also administered the 25-item multiple-choice test as a pre- and
post-test 
to students enrolled in a sophomore-junior level E\&M course in which vector
calculus was used extensively; 33 students took the pretest and 28 students
took the posttest. Tables~\ref{tab3} and \ref{tab4} show the percentage of students who made the various choices
on various questions on the pretest and
posttest respectively. The average pre- and posttest scores are
44\% and 49\% respectively; this difference is not statistically
significant. The fact that the upper-level undergraduates did not perform
better than the introductory students (compare Tables~\ref{tab1} and \ref{tab4}) suggests that
the higher mathematical sophistication of the course with vector
calculus did not help students acquire a better conceptual
understanding of the superposition principle, symmetry, and Gauss's law.

\subsection{Performance of Graduate Students}

To calibrate the test we also administered it (over two consecutive years)
to a total of 33 physics graduate students who were enrolled in a seminar
course for teaching assistants (TAs). Most of them were first year
graduate students who were simultaneously enrolled in the first semester of
the graduate E\&M course. Students were told ahead of time that they would
be taking a test related to electrostatics concepts. They were asked to take
the test seriously, but it did not count for their course grade. The average
test score for the graduate students was approximately 75\% with the
reliability coefficient $\alpha
\approx 0.8$.\cite{nitko} The better performance of graduate students
compared to the undergraduates is statistically significant. The minimum
score obtained by a graduate student (an American student) was 28\% and
the maximum score obtained by two Chinese graduate students was 100\%.

Table~\ref{tab5} shows the percentage of graduate students who selected the various choices
on the test. Many of the conceptual
difficulties that the graduate students displayed are similar to those of
the introductory students. \prob s~13 and 17 were the only ones on which
graduate students performed less than 50\% with misconceptions similar to
those of the undergraduates. The PBD for \prob s~13 and 17 were
$0.5$ and $0.6$ respectively, which shows that the graduate students who
performed well overall on the test did well on those questions. Comparison
of Tables~\ref{tab1} and \ref{tab5} shows that on an average, the graduate students
outperformed the introductory students on all other questions except
\prob~2, but this difference is not statistically significant. The PBD for this question is $0.5$, so the graduate students who performed
well overall on the test did well on this question.

\section{Summary}

We find that the undergraduate students,
including those in the upper-level course with higher mathematical
sophistication, have common difficulties related to the superposition
principle, symmetry concepts, and Gauss's law. The knowledge deficiencies can
be broadly divided into three levels with increasing difficulty: lack of knowledge related to a particular concept, knowledge
that is retrieved from memory but cannot be interpreted correctly, and
knowledge that is retrieved and interpreted at the basic level but cannot be
used to draw inferences in specific situations. Our investigation shows
evidence that students' difficulties were due to knowledge deficiencies
across all three levels. Because many questions required students to
predict the outcomes for specific setups, they necessitated a
transition from a mathematical representation to a concrete case. 
Therefore, deficiencies at the latter two levels were frequently observed. 
Instructional strategies that focus on improving student understanding of
these concepts should take into account these difficulties. The
multiple-choice test that we developed can help assess the
effectiveness of strategies to improve student understanding of these
concepts.

\begin{acknowledgments}
We are very grateful to P.\ Reilly and all the faculty 
who reviewed the various components of the test at various stages and
provided invaluable feedback. We are also very thankful to all the faculty
who administered the test. We thank R.\ Devaty, A.\ Janis, R.\ Johnsen, and
J.\ Levy for a critical reading of the manuscript. This work is supported in part
by the National Science Foundation award DUE-0442087.
\end{acknowledgments}

\section{\label{app:tables}Tables of results}

\begin{table}[h!]
\begin{tabular}[t]{|c|c|c|c|c|c|c|c|c|c|c|c|c|c|c|c|c|c|c|c|c|c|c|c|c|c|}
\hline
Q & 1 & 2& 3&4&5&6&7&8&9&10&11&12&13&14&15&16&17&18&19&20&21&22&23&24&25 \\
\hline 
\hline
(a) &\bf{\it{55}} &3&20&0&14&3&5&4&7&20&4&\bf{\it{57}}&55&12&40&4&0&\bf{\it{47}}&2&22&28&\bf{\it{37}}&35&10&\bf{\it{32}}\\
\hline
(b) &35&13&\it{57}&30&0&11&6&\it{50}&\it{66}&0&5&6&10&24&8&10&\it{33}&7&3&14&5&15&2&15&33\\
\hline
(c) &5&\it{69}&15&5&0&8&\it{81}&5&12&19&\it{57}&30&8&2&\it{43}&18&25&27&34&6&17&13&\it{53}&27&22\\
\hline
(d) &1&7&7&15&0&\it{58}&6&35&5&\it{50}&6&7&6&\it{60}&4&42&17&13&\it{56}&\it{53}&\it{32}&17&3&19&12\\
\hline
(e) & 4&8&1&\it{50}&\it{86}&20&2&6&10&11&28&0&\it{21}&2&5&\it{26}&25&6&5&5&18&18&7&\it{29}&1\\
\hline
\end{tabular}
\caption{\label{tab1}Percentage of introductory calculus-based physics
students (total number of students 541) who selected choices (a)--(e) on
\prob s~(1)--(25) on the test. The correct response for each question has
been italicized. The average score was 49\%.}
\end{table}

\begin{table}[h!]
\begin{tabular}{|p{9.0cm}|p{5.5cm}|}
\hline
\hline
{\bf Concepts} &
{\bf \prob\ number} \\
\hline
\hline
Electric flux& 1, 7, 8, 9, 12, 18, 20, 21, 23 \\
\hline
Recognizing the symmetry of the charge distribution & 2, 3, 4, 6, 10, 11,
12, 13, 14, 15, 16, 17, 22, 24, 25 \\
\hline
Symmetry of the object versus symmetry of the charge distribution & 6, 10, 14,
15, 16\\
\hline
Coulomb's law, superposition and symmetry considerations sufficient & 2, 3,
4, 14, 16\\
\hline
Difference between the electric field and the electric flux& 1, 7, 8, 12, 13, 15, 18, 19, (21,\,22), 23\\
\hline
Relevance of a closed surface in Gauss's law& 5, 21, 22\\
\hline
Recognizing the symmetry to determine if it is easy to exploit Gauss's law
or exploiting Gauss's law to determine the electric field& 6, 10, 11, 12,
13, 14, 15, 16, 17, 22, 24, 25\\
\hline
Appropriate Gaussian surface for determining the electric field for a given
charge distribution & 6, 11, 22\\
\hline
Electric field inside hollow non-conducting objects with different charge distribution &13, 15, 25\\
\hline
\hline
\end{tabular}
\caption{\label{tab2}Concepts covered and the questions that addressed them in the multiple-choice test.}
\end{table}

\begin{table}[h!]
\begin{tabular}[t]{|c|c|c|c|c|c|c|c|c|c|c|c|c|c|c|c|c|c|c|c|c|c|c|c|c|c|}
\hline
Q & 1 & 2& 3&4&5&6&7&8&9&10&11&12&13&14&15&16&17&18&19&20&21&22&23&24&25 \\
\hline 
\hline
(a) &\bf{\it{42}} &9&15&3&30&0&9&0&9&24&3&\bf{\it{43}}&49&9&49&9&0&\bf{\it{58}}&0&15&15&\bf{\it{27}}&42&12&\bf{\it{27}}\\
\hline
(b) &49&15&\it{61}&36&0&3&18&\it{52}&\it{67}&0&15&18&15&24&3&21&\it{27}&3&6&18&13&15&3&27&30\\
\hline
(c) &6&\it{52}&9&9&0&3&\it{64}&12&15&21&\it{24}&33&15&9&\it{36}&27&30&9&24&9&15&12&\it{42}&30&21\\
\hline
(d) &3&21&15&18&0&\it{82}&0&24&6&\it{45}&18&3&3&\it{55}&3&33&6&18&\it{64}&\it{55}&\it{36}&30&3&12&21\\
\hline
(e) & 0&3&0&\it{33}&\it{70}&12&9&12&3&9&39&3&\it{18}&3&9&\it{9}&36&12&6&3&21&15&9&\it{18}&0\\
\hline
\end{tabular}
\caption{\label{tab3}Percentage of students in the upper-level undergraduate E\& M course (total number of students 33)
who selected choices (a)--(e) on the pretest (before instruction in the upper-level course). 
The correct response for each question is italicized. The average score was 44\%.}
\end{table}

\begin{table}[h!]
\begin{tabular}[t]{|c|c|c|c|c|c|c|c|c|c|c|c|c|c|c|c|c|c|c|c|c|c|c|c|c|c|}
\hline
Q & 1 & 2& 3&4&5&6&7&8&9&10&11&12&13&14&15&16&17&18&19&20&21&22&23&24&25 \\
\hline 
\hline
(a) &\bf{\it{53}} &11&11&4&32&0&11&11&4&21&0&\bf{\it{53}}&75&0&64&4&0&\bf{\it{39}}&0&7&7&\bf{\it{50}}&36&4&\bf{\it{28}}\\
\hline
(b) &39&29&\it{61}&46&0&11&11&\it{32}&\it{68}&4&4&11&0&32&4&7&\it{7}&7&4&21&4&4&4&7&32\\
\hline
(c) &0&\it{57}&11&7&0&4&\it{68}&7&18&11&\it{68}&25&4&4&\it{25}&7&43&32&36&14&32&18&\it{53}&18&0\\
\hline
(d) &4&0&14&4&0&\it{82}&11&36&7&\it{50}&0&7&7&\it{64}&7&46&11&11&\it{53}&\it{53}&\it{53}&14&0&11&36\\
\hline
(e) & 4&4&4&\it{39}&\it{68}&4&0&14&4&14&28&4&\it{14}&0&0&\it{36}&39&11&7&4&4&14&7&\it{61}&4\\
\hline
\end{tabular}
\caption{\label{tab4}Percentage of students in the upper-level undergraduate E\& M course (total number of students 28)
who selected the choices (a)--(e) on 
the posttest (after instruction in the upper-level course). 
The correct response for each question is italicized. The average score was 49\%.} 
\end{table}

\begin{table}[h!]
\begin{tabular}[t]{|c|c|c|c|c|c|c|c|c|c|c|c|c|c|c|c|c|c|c|c|c|c|c|c|c|c|}
\hline
Q & 1 & 2& 3&4&5&6&7&8&9&10&11&12&13&14&15&16&17&18&19&20&21&22&23&24&25 \\
\hline 
\hline
(a) &\bf{\it{73}} &3&12&0&12&3&3&6&0&12&0&\bf{\it{79}}&55&0&36&6&0&\bf{\it{73}}&3&3&6&\bf{\it{73}}&18&0&\bf{\it{55}}\\
\hline
(b) &27&9&\it{79}&18&0&3&3&\it{88}&\it{91}&0&3&0&0&12&3&6&\it{45}&0&0&3&3&3&0&6&9\\
\hline
(c) &0&\it{64}&9&0&0&6&\it{91}&0&0&6&\it{94}&12&9&3&\it{61}&9&40&24&0&0&0&12&\it{82}&9&12\\
\hline
(d) &0&3&0&9&0&\it{76}&0&6&6&\it{70}&0&6&0&\it{85}&0&27&0&0&\it{94}&\it{94}&\it{88}&12&0&30&24\\
\hline
(e) & 0&21&0&\it{73}&\it{88}
&12&3&0&3&12&3&3&\it{36}&0&0 & \it{52}&15&3&3&0&3&0&0&\it{55}&0\\
\hline
\end{tabular}
\caption{\label{tab5}Percentage of physics graduate students enrolled in a
course for teaching assistants (total number of students 33) who selected
the choices (a)--(e) on Problems (1)--(25) on the test. The correct
response for each question has been italicized. The average score was
$75\%$.}
\end{table}

\clearpage

\appendix

\section{\label{app:summary}Summary of the Multiple-choice Test Design}

During the design of the multiple-choice test,
we paid particular attention to the issues of reliability and
validity.\cite{nitko} Reliability refers to the relative degree of
consistency between testing if the test procedures are repeated for an
individual or group. Validity refers to the appropriateness of the test
score interpretation. The test design began with the development of a test
blueprint that provided a comprehensive framework for planning decisions
about the desired test attributes. The degree of specificity in the test
plan was useful for creating questions. We tabulated the scope and extent of the
content covered and the level of cognitive complexity desired. We used
previous free-response and multiple-choice questions administered to
students as a guide and identified the desired performance and a
description of conditions/contexts under which the performance was expected
to occur. 

We classified the cognitive complexity using a simplified version of
Bloom's taxonomy\cite{bloom}: specification of knowledge, interpretation of
knowledge, and application of knowledge in different situations. The
performance targets and table of content and cognitive complexity were shown
to five physics faculty members at the University of Pittsburgh for
review. Modifications were made to the weights assigned to various concepts
and to the performance targets based on faculty feedback. The
performance targets were converted to approximately 30 free-response
questions. These questions required students to provide reasoning for their
responses. The free-response questions were administered (in groups of
10--20) to students in the calculus-based courses. Often, some
students in a class were given one set of questions and others were given
another set in order to sample student responses on most of the questions.
We also tape-recorded interviews with 5 introductory student volunteers
using a think-aloud protocol.\cite{chi} 30 multiple-choice questions were
then designed using the most frequent incorrect student responses for the
free-response questions and interviews as a guide for making the
distractor choices. Choosing the four distractors to conform to
the common difficulties is essential for increasing the discriminating
properties of the questions. Five physics faculty members were asked to
review the multiple-choice questions and comment on their appropriateness
and relevance for calculus-based introductory physics and to detect
ambiguity in question wording. A review form was developed to aid the
faculty in reviewing the questions. The faculty also classified each
question on a scale from very appropriate to least appropriate. Further
modifications were made based on their recommendations. Then, a
multiple-choice test was assembled using 25 questions that closely matched
the initial table delineating the scope of the content and cognitive
complexity. The same faculty members who earlier reviewed the questions were
shown the test several times and modifications were made based on the
feedback during the iterations.

The 50 minute multiple-choice test was administered after instruction in
Gauss's law to students in the calculus-based courses at Pittsburgh. Five
student volunteers who had not taken the test earlier were interviewed
individually and asked to respond to the test questions using the
think-aloud protocol. These interviews provided us with an opportunity to
clarify issues in-depth. The reliability index $\alpha$ for the test was
approximately 0.8 which is good by the standards of
test-design.\cite{nitko} Item analysis of student responses was performed to
judge whether each question functioned as expected.\cite{nitko} In addition
to the calculation of difficulty and discrimination of
questions,\cite{nitko} item analysis included creating a table to count the
number of students selecting each distractor in the upper and lower
quartiles. Item analysis was very useful to determine whether individual
questions and distractors functioned as expected. Based upon the item
analysis and interviews, the test questions were modified further before
being administered to the students.

\clearpage
\section{\label{app:test}The multiple-choice test}
\begin{center}
\begin{boxedminipage}{18cm}
{\bf Instructions}: {Select \emph{one} of the five choices (a)--(e) for each
of the 25 \prob s.} In all questions
all physical objects are insulating (non-conducting);
all insulators are nonpolarizable;
$\epsilon_0$ is the permittivity of free space;
Gauss's law: $\oint \vert \vec E \vert \vert \vec {dA} \vert \cos
\theta =Q_{\rm enc}/\epsilon_0$; and 
the sign convention is that for all closed surfaces, consider outward flux
as positive.
\end{boxedminipage}
\end{center}

\begin{enumerate}[Problem 1.]

\item Choose {\em all} of the following physical variables that are vectors:
\vspace*{-.15in}
\begin{enumerate}[(i)]
\setlength{\itemsep}{-0.1ex plus 0.1ex minus 0.1ex}
\setlength{\itemindent}{1cm}
\item Electric field \label{temp}
\item Electric flux
\item Electric charge
\end{enumerate}

\vspace{-0.5cm}
\begin{enumerate}[(a)]
\setlength{\itemsep}{-0.1ex plus 0.1ex minus 0.1ex}
\item (i) only.
\item (i) and (ii) only.
\item (i) and (iii) only.
\item (ii) and (iii) only.
\item (i), (ii), and (iii).
\end{enumerate}

\item \label{problem2}Three identical point charges $+Q$ are arranged in a
line as shown in Fig.~\ref{figprob2}. Points A, B, and C are along a
parallel line. You do {\it not} know the lengths $L$ and $d$. The three
charges produce an electric field. Without knowledge of $L$ and $d$, what
can you infer about the electric field at points A, B, and C?
\vspace*{-.15in}
\begin{center}
\begin{figure}[h!]
\includegraphics[width=2.2in]{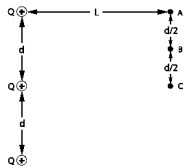}
\caption{\label{figprob2} Charge distribution for Problem~\ref{problem2}.}
\end{figure}
\vspace*{-.15in}
\end{center}

\begin{enumerate}[(a)]
\vspace{-1cm}
\setlength{\itemsep}{-0.1ex plus 0.1ex minus 0.1ex}
\item Both the magnitude and direction of the field are the same at points A,
B, and C.
\item The magnitude of the field is the same at points A, B, and C but
the directions are different.
\item The exact direction of the field can be
predicted only at point C.
\item The exact direction of the field can be
predicted only at points A and C.
\item The exact direction of the field
cannot be predicted at any of the three points.
\end{enumerate}

\item \label{prob3}Consider a horizontal square sheet, length $L$ on each side, on which
positive charge is uniformly distributed with charge per unit area (surface
charge density $\sigma$\,C/m$^2$) (see Fig.~\ref{finitesheet}). You measure the electric field at two
points, each at a height $h=L/2$ above the sheet: point C is directly 
above the center of the sheet and point B is off center. 

\vspace*{-.18in}
\begin{center}
\begin{figure}[h!]
\includegraphics[width=2.2in]{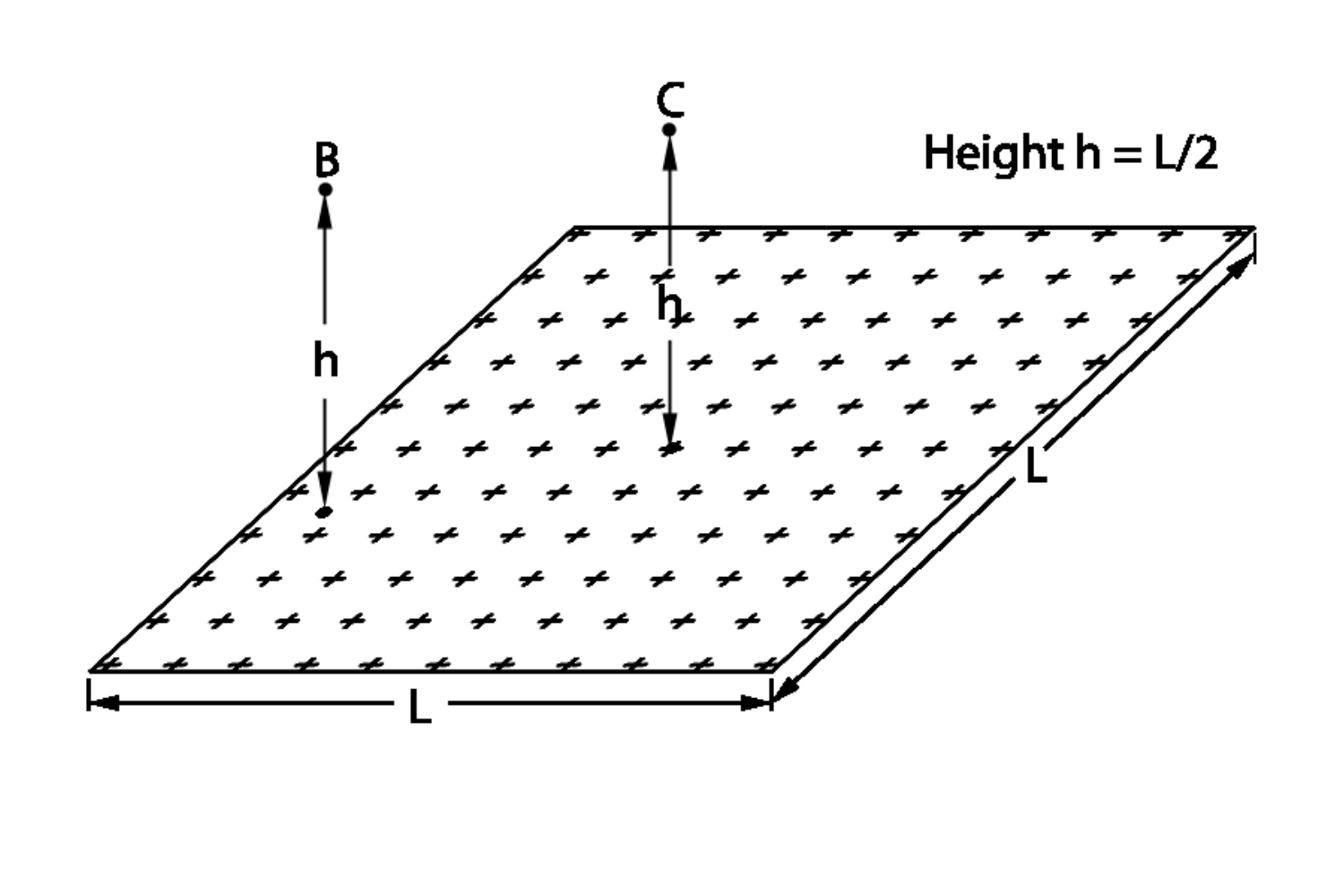}
\vspace{-.6cm}
\caption{\label{finitesheet} A horizontal square sheet of charge for Problem~\ref{prob3}.}
\end{figure}
\end{center}
\vspace*{-.25in}

Which one of the following statements is true about the 
field due to the finite sheet of charge, observed at points B and C?
\begin{enumerate}[(a)]
\setlength{\itemsep}{-0.1ex plus 0.1ex minus 0.1ex}
\item The fields at points B and C have the same magnitude and same
direction.
\item The fields at points B and C have different magnitudes and
different directions.
\item The fields at points B and C have the same
magnitude but different directions.
\item The fields at points B and C have
different magnitudes but the same direction.
\item We cannot compare the
fields at points B and C without knowing the numerical value of $\sigma$.
\end{enumerate}

\item \label{problem4}You perform two experiments (E1 and E2) in which you
distribute charge $+3Q$ differently on an equilateral triangle made with
thin insulating rods. E1: You put identical charges, $+Q$ each, in three
localized blobs on the vertices of the triangle.
E2: You distribute
charge $+3Q$ uniformly on the triangle. 
The dashed triangle in Fig.~\ref{figprob4} shows an imaginary equilateral
triangle concentric with the insulating triangle. Which one of the following statements is true about the electric
field magnitude at points on the dashed imaginary
triangle due to the $+3Q$ charge?
\vspace*{-.25in}
\begin{center}
\begin{figure}[h]
\includegraphics[width=2.2in]{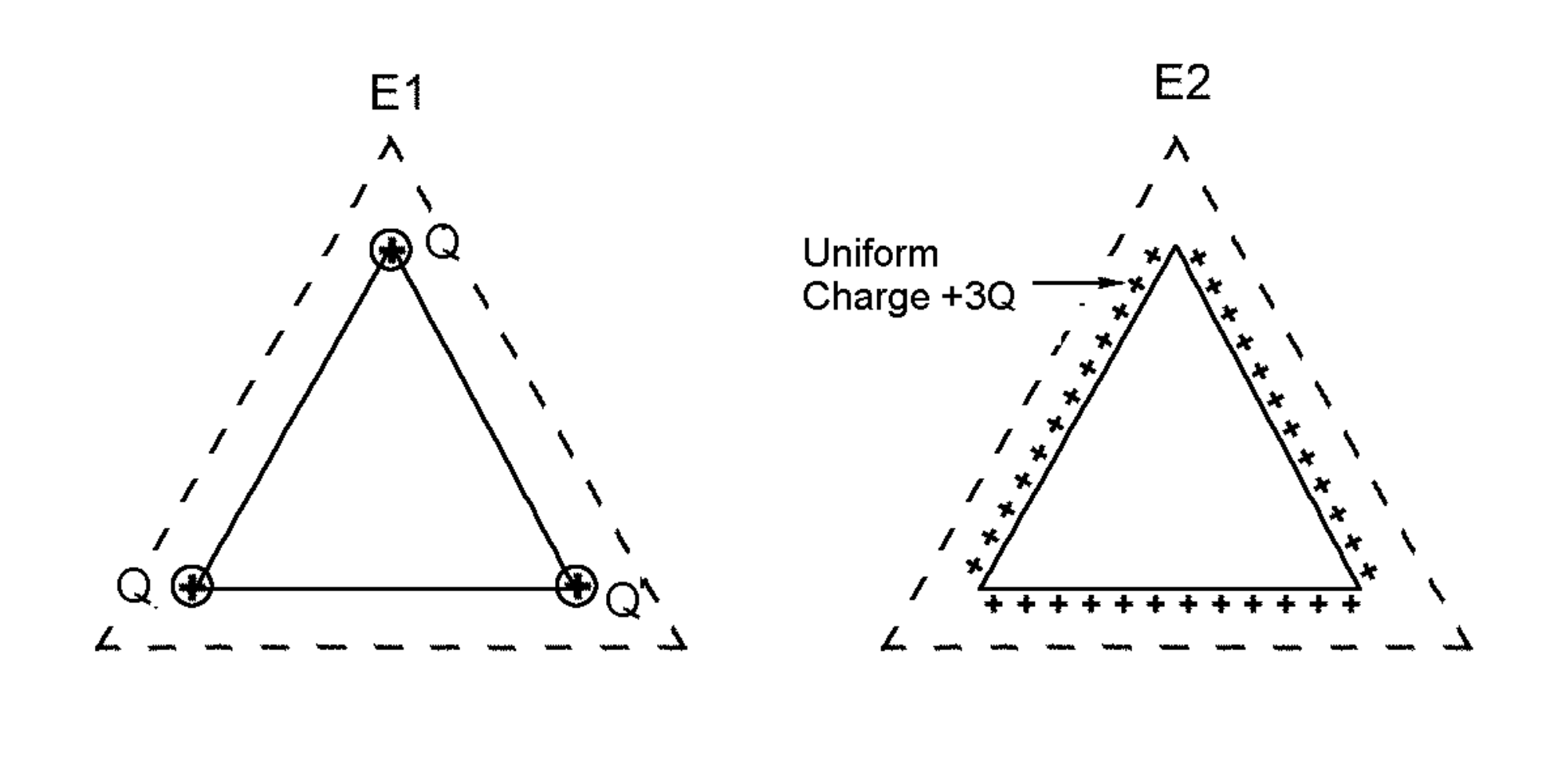}
\vspace{-.6cm}
\caption{\label{figprob4}Charge distribution for Problem~\ref{problem4}.}
\end{figure}
\end{center}

\begin{enumerate}[(a)]
\setlength{\itemsep}{-0.1ex plus 0.1ex minus 0.1ex}

\item It is the same everywhere on the dashed triangle only in experiment
E1.

\item It is the same everywhere on the dashed triangle only in experiment
E2.

\item In each experiment, it is the same everywhere on the dashed triangle,
but the magnitudes differ in the two experiments.

\item In each experiment, it is the same everywhere on the dashed triangle,
and the magnitudes are equal in the two experiments.

\item Both in experiment E1 and in experiment E2 it varies from point to
point on the dashed triangle.
\end{enumerate}

\item
For Gauss's law to be valid, the Gaussian surface used {\it must} be a
\begin{enumerate}[(a)]
\setlength{\itemsep}{-0.1ex plus 0.1ex minus 0.1ex}
\item highly symmetrical surface.
\item spherical surface.
\item cylindrical surface.
\item open surface.
\item closed surface.
\end{enumerate}

\item
Choose all of the following statements that must be true about a Gaussian
surface in order for Gauss's law to be convenient for calculating the
electric field at a point on the surface:

\vspace{-0.5cm}

\begin{enumerate}[(i)]
\setlength{\itemsep}{-0.1ex plus 0.1ex minus 0.1ex}
\setlength{\itemindent}{1cm}

\item The electric field direction should be easy to predict at every point
on the surface.

\item The Gaussian surface must be chosen to take advantage of the symmetry
of the charge distribution.

\item The Gaussian surface must be chosen to take advantage of the symmetry
of the object enclosed inside it regardless of how the charges are
distributed on that object.
\end{enumerate}
\begin{enumerate}[(a)]
\setlength{\itemsep}{-0.1ex plus 0.1ex minus 0.1ex}
\item (i) only.
\item (ii) only.
\item (iii) only.
\item (i) and (ii) only.
\item (i) and (iii) only.
\end{enumerate}

\item \label{problem7} In Fig.~\ref{probfig7}, a point charge $+Q_1$ is at the
center of an imaginary spherical surface and another point charge $+Q_2$
is outside it. Point P is on the surface of the sphere. Let $\Phi_S$ be the net electric flux through the sphere and $\vec E_P$
be the electric field at point P on the sphere. Which one of the following statements is true?

\begin{center}
\begin{figure}[h]
\includegraphics[width=2.2in]{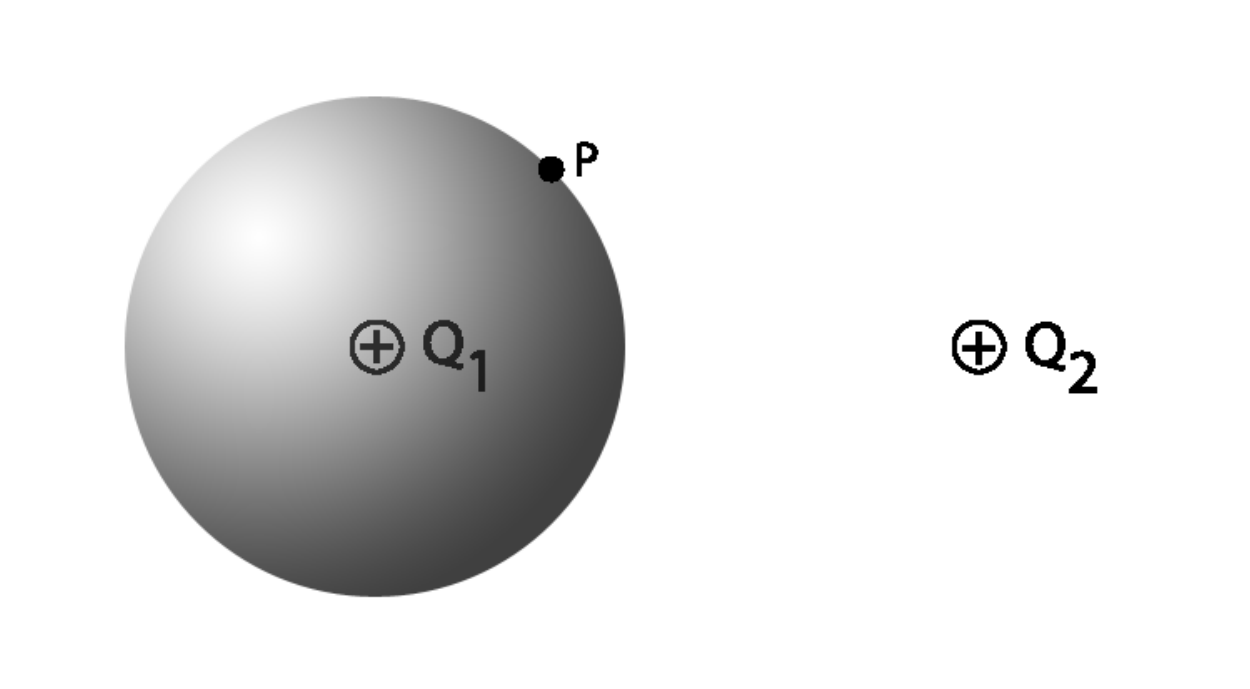}
\caption{\label{probfig7}Charge distribution for Problem~\ref{problem7}.}
\end{figure}
\end{center}

\begin{enumerate}[(a)]
\setlength{\itemsep}{-0.1ex plus 0.1ex minus 0.1ex}
\item Both charges $+Q_1$ and $+Q_2$ make nonzero contributions to $\Phi_S$
but only the charge $+Q_1$ makes a nonzero contribution to $\vec E_P$. 
\item Both charges $+Q_1$ and $+Q_2$ make nonzero contributions to $\Phi_S$
but only the charge $+Q_2$ makes a nonzero contribution to $\vec E_P$. 
\item Only the charge $+Q_1$ makes a nonzero contribution to $\Phi_S$ but
both charges $+Q_1$ and $+Q_2$ make nonzero contributions to $\vec E_P$. 
\item Charge $+Q_1$ makes no contribution to $\Phi_S$ or $\vec E_P$.
\item Charge
$+Q_2$ makes no contribution to $\Phi_S$ or $\vec E_P$.
\end{enumerate}

\item
Your friend measures the electric flux through three closed surfaces (1), (2), and (3) to be 1\,Nm$^2$/C, 2\,Nm$^2$/C, and $-3$\,Nm$^2$/C respectively.
Choose all of the following statements that can be inferred from these measurements:

\vspace*{-.15in}
\begin{enumerate}[(i)]
\setlength{\itemsep}{-0.1ex plus 0.1ex minus 0.1ex}
\setlength{\itemindent}{1cm}

\item The area of surface (3) is largest.

\item The magnitude of the net charge enclosed inside surface (3) is largest.
\item The electric field everywhere on surface (1) is weaker than on surface (2).
\end{enumerate}
\begin{enumerate}[(a)]
\setlength{\itemsep}{-0.1ex plus 0.1ex minus 0.1ex}
\item (i) only.
\item (ii) only.
\item (i) and (ii) only.
\item (ii) and (iii) only.
\item (i), (ii), and (iii).
\end{enumerate}

\item \label{prob9} Shown in Fig.~\ref{concentric} are three concentric spherical Gaussian surfaces A, B, and C with a positive point charge $+Q$ at their center.
A second, but negative point charge $-Q$ is enclosed only by surface C. 
\vspace*{-.22in}
\begin{center}
\begin{figure}[h]
\includegraphics[width=2.2in]{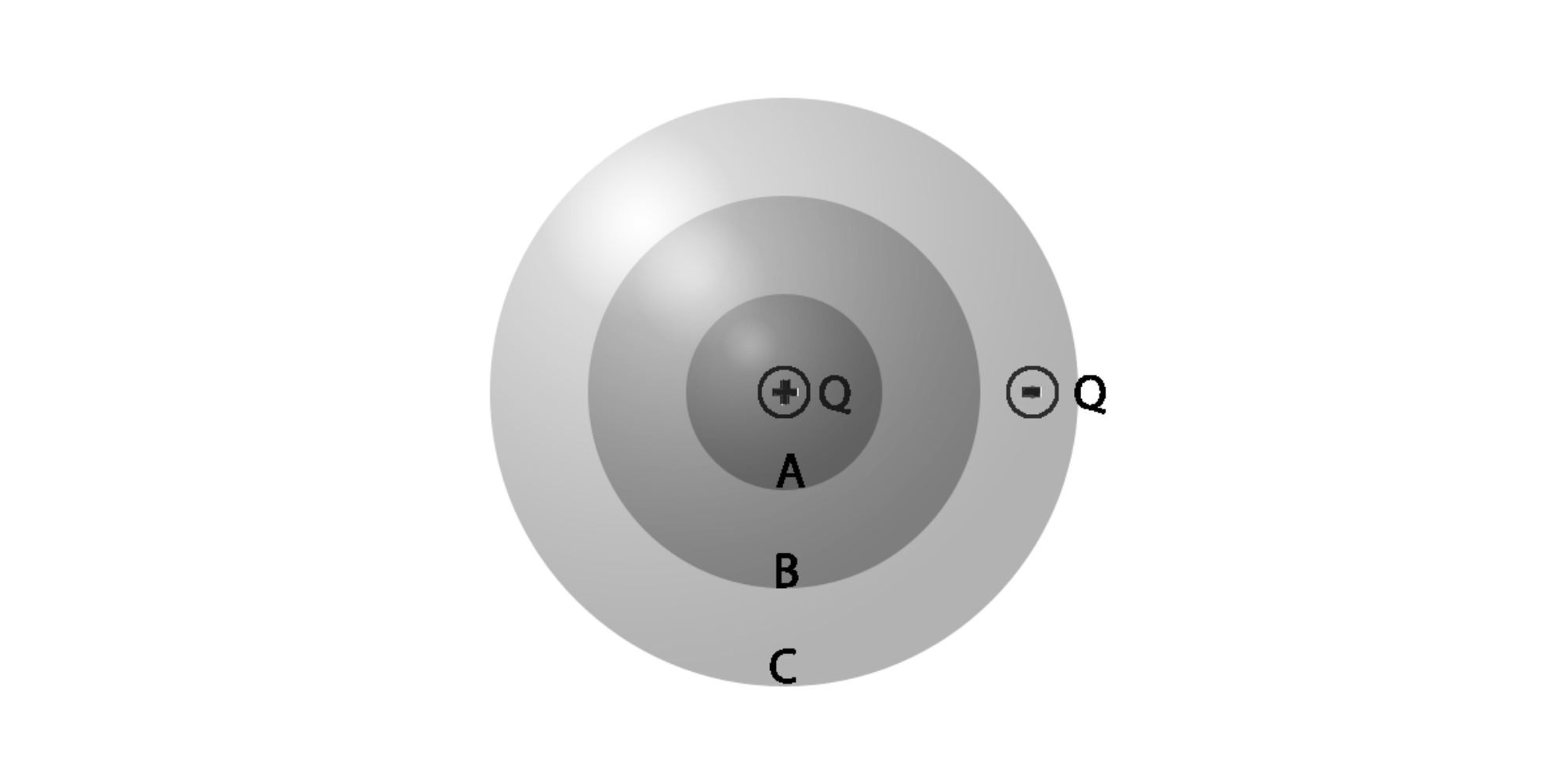}
\vspace{-0.5cm}
\caption{\label{concentric}Gaussian surfaces for Problem~\ref{prob9}.}
\end{figure}
\end{center}
\vspace*{-.4in}
Which is a correct statement about the magnitudes of the electric flux
$\Phi_S$ through the three surfaces?
\begin{enumerate}[(a)]
\setlength{\itemsep}{-0.1ex plus 0.1ex minus 0.1ex}
\item $\Phi_A=\Phi_B=\Phi_C$.
\item $\Phi_A=\Phi_B>\Phi_C$.
\item $\Phi_A>\Phi_B>\Phi_C$.
\item $\Phi_B>\Phi_A>\Phi_C$.
\item None of the above.
\end{enumerate}

\item \label{prob10}Choose all of the following cases for which the electric field at any
point outside the object can be calculated easily from Gauss's law. (In each
case, assume that the insulators are nonpolarizable and no other charges are present
anywhere.)

\vspace*{-.15in}
\begin{enumerate}[(i)]
\setlength{\itemsep}{-0.1ex plus 0.1ex minus 0.1ex}
\setlength{\itemindent}{1cm}

\item Insulating sphere with a uniform charge throughout its volume.
\item Insulating dumbbell with a uniform charge throughout its volume.

\item Insulating dumbbell with only one of the two spherical balls at the end uniformly charged throughout its volume.
\end{enumerate}
\begin{center}
\begin{figure}[h]
\includegraphics[width=1.0in]{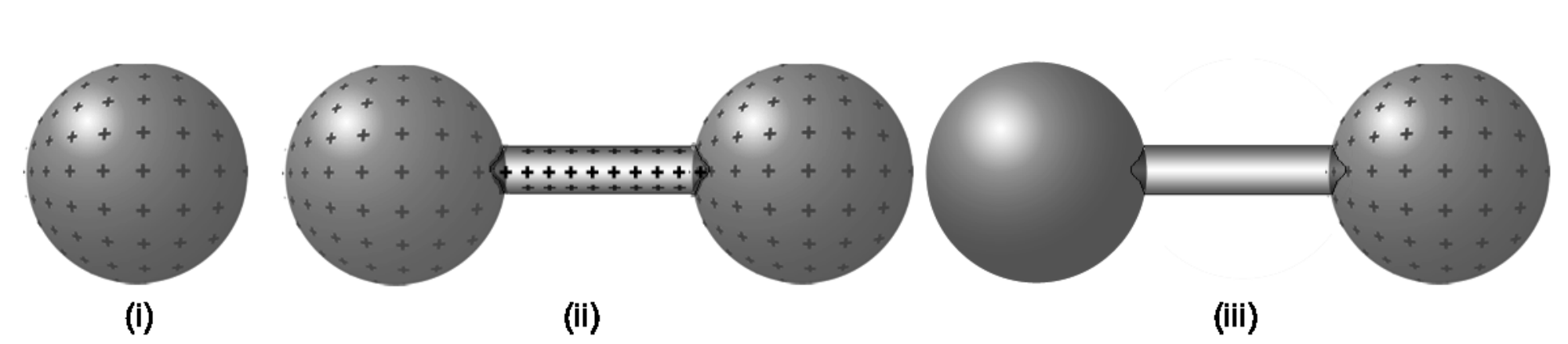}
\caption{\label{dumb} Diagram for Problem~\ref{prob10}.}
\end{figure}
\end{center}
\vspace{-0.5cm}
\begin{enumerate}[(a)]
\setlength{\itemsep}{-0.1ex plus 0.1ex minus 0.1ex}
\item (i) only.
\item (ii) only.
\item (i) and (ii) only.
\item (i) and (iii) only.
\item (i), (ii), and (iii).
\end{enumerate}

\item \label{prob11}Consider three possible Gaussian surfaces (a sphere, a cube, and a
cylinder) which extend half above and half below an infinite horizontal
sheet of uniform charge density as shown below in Fig.~\ref{three}.
\vspace*{-.11in}
\begin{center}
\begin{figure}[h]
\includegraphics[width=2.09in]{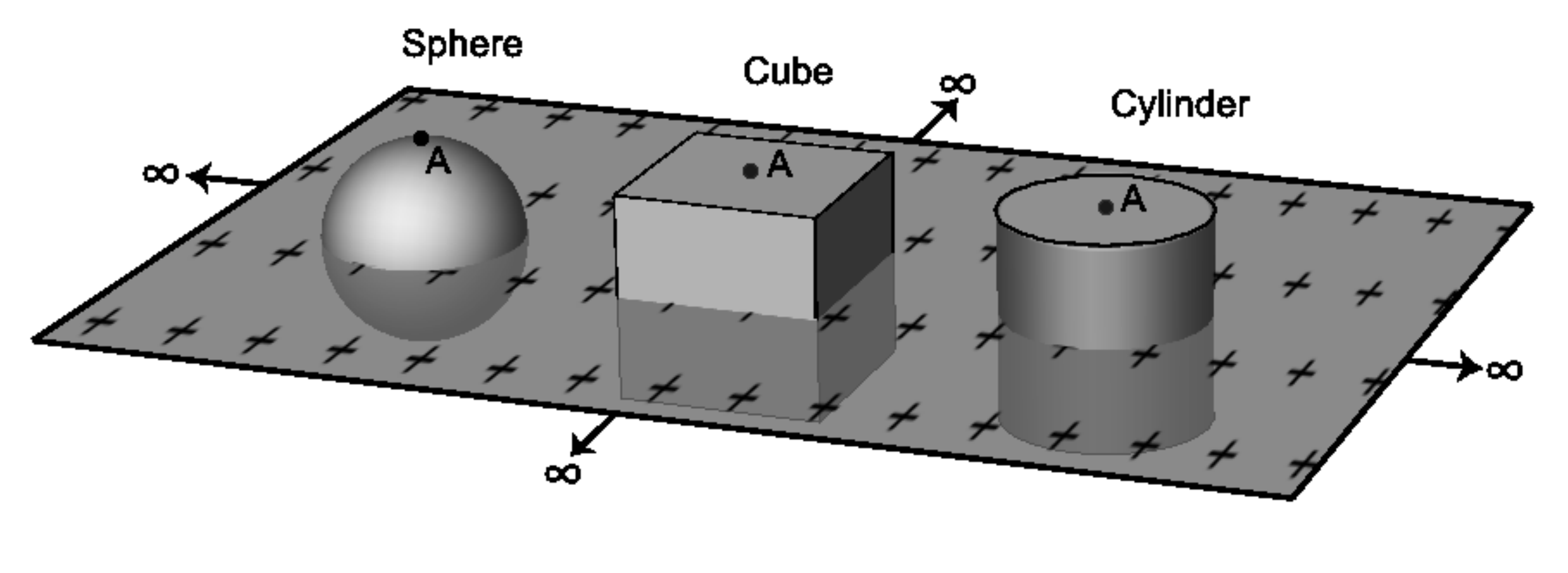}
\caption{\label{three} Diagram for Problem~\ref{prob11}.}
\end{figure}
\end{center}
\vspace*{-.2in}
Point A is located at the top center of each Gaussian surface.
For which of the Gaussian surfaces will Gauss's law help us to easily calculate the electric field at point A due to the sheet of charge?
\begin{enumerate}[(a)]
\setlength{\itemsep}{-0.1ex plus 0.1ex minus 0.1ex}

\item Only the sphere is symmetric enough.
\item Only the cylinder, because the side walls have zero flux and it has
circular symmetry.
\item Only the cylinder and the cube, because any shape
with the side walls perpendicular to the sheet and end caps parallel to the
sheet will work.
\item Only the sphere and the cylinder, because they have
circular cross section.
\item All surfaces will work since they are
symmetric.
\end{enumerate}

\item \label{prob12} A thin insulating rod of length 1\,m, with charge $Q=+100$\,nC (nanocoulombs) uniformly distributed over it, is symmetrically
situated inside a spherical Gaussian surface with a total surface area of $A=15$\,m$^2$ (see Fig.~\ref{four}). No other charges are present anywhere. 
\vspace*{-.10in}
\begin{center}
\begin{figure}[h]
\includegraphics[width=1.8in]{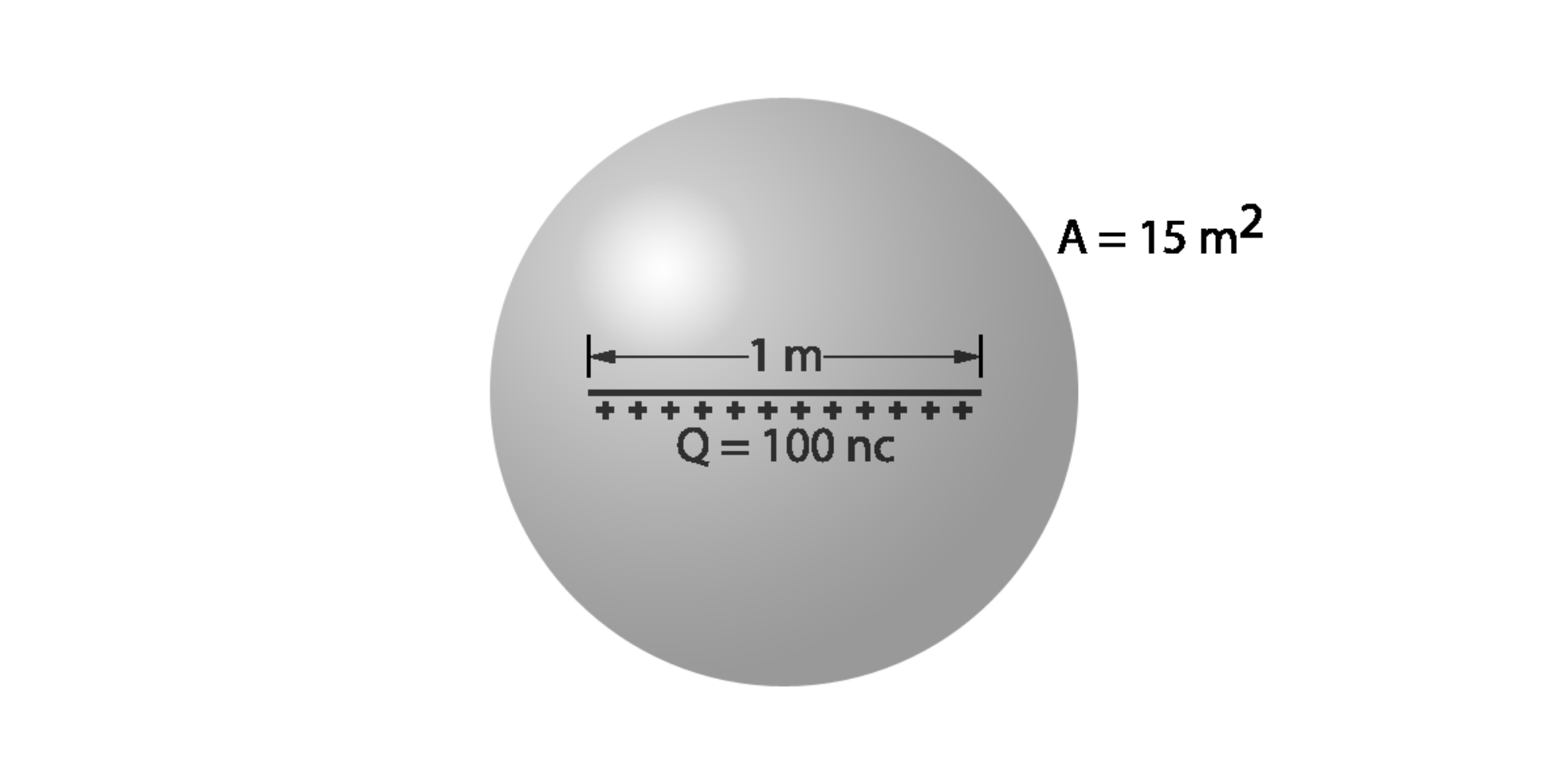}
\caption{\label{four} Diagram for Problem~\ref{prob12}.}
\end{figure}
\end{center}
\vspace*{-.2in}
We can use Gauss's law to conclude that:
\vspace*{-.15in}
\begin{enumerate}[(i)]
\setlength{\itemsep}{-0.1ex plus 0.1ex minus 0.1ex}
\setlength{\itemindent}{1cm}
\item the magnitude of the net electric flux through the Gaussian surface is $\Phi_S=Q/\epsilon_0$.
\item the electric field magnitude at any point on the surface is $\vert \vec E \vert=\Phi_S/A=Q/(\epsilon_0 A)$.
\end{enumerate}
Which of these statements is true?
\begin{enumerate}[(a)]
\setlength{\itemsep}{-0.1ex plus 0.1ex minus 0.1ex}
\item (i) only.
\item (ii) only.
\item both (i) and (ii).
\item neither (i) nor (ii).
\item Not enough information.
\end{enumerate}

\item
The surface of a thin-walled cubic insulating (non-conducting) box is given
a uniformly distributed positive surface charge. Which one of the following
can be inferred about the electric field everywhere inside the 
insulating box due to this surface charge using Gauss's law?
\begin{enumerate}[(a)]
\setlength{\itemsep}{-0.1ex plus 0.1ex minus 0.1ex}
\item Its magnitude everywhere inside must be zero.
\item Its magnitude everywhere inside must be nonzero but uniform (the
same).
\item Its direction everywhere inside must be radially outward from
the center of the box.
\item Its direction everywhere inside must be
perpendicular to one of the sides. 
\item None of the above.
\end{enumerate}

\end{enumerate}

\begin{center}
\begin{figure}[h]
\includegraphics[width=1.8in]{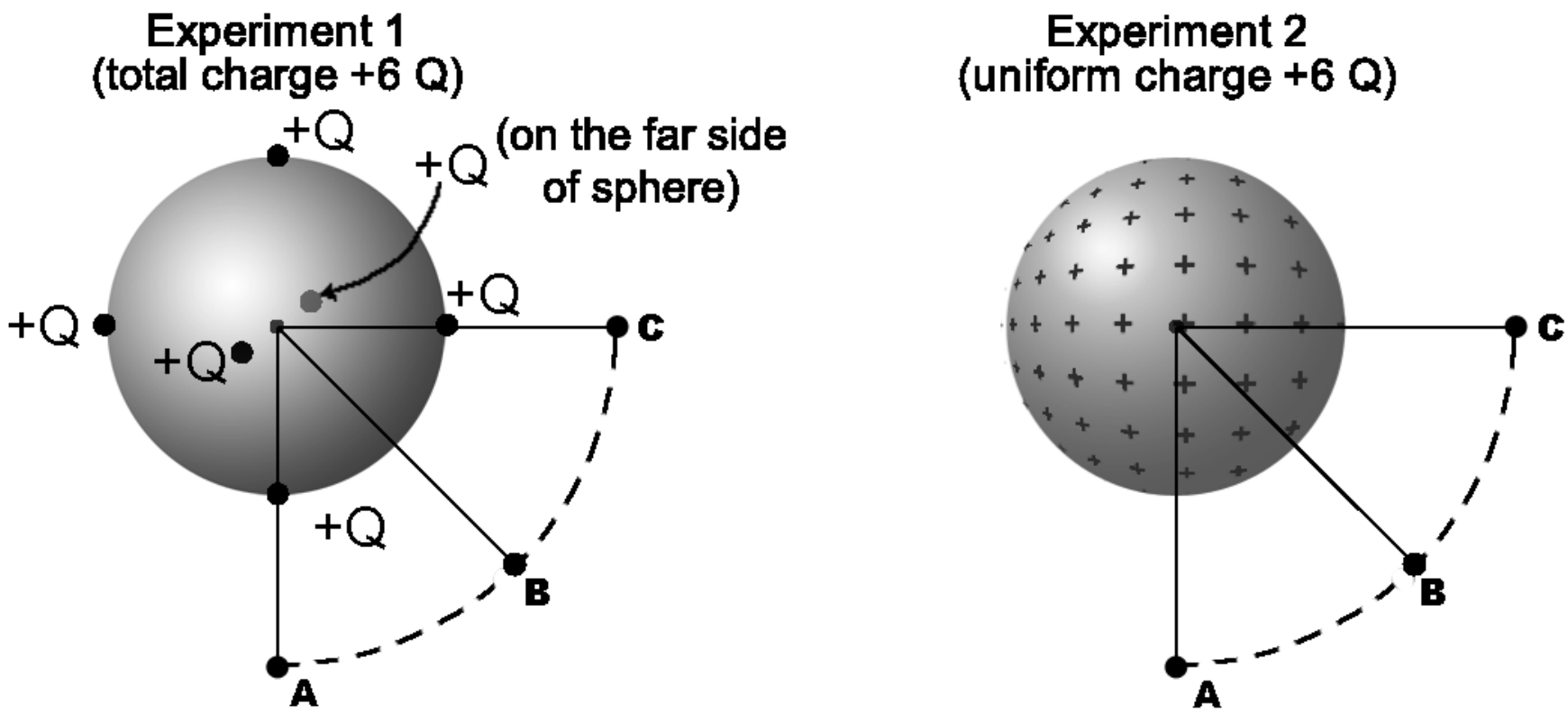}
\caption{\label{fig14}Diagram for Problems~\ref{problem14} and
\ref{problem15}.}
\end{figure}
\end{center}

\noindent {\bf Setup for Problems~\ref{problem14} and \ref{problem15}}\\
You perform two experiments (see Fig.~\ref{fig14}) in which you distribute
charge
$+6Q$ differently on the surface of an isolated hollow insulating
(non-conducting) sphere: Experiment 1: You put identical charges, $+Q$ each, on
the spherical surface in six localized blobs (you can consider them point
charges) such that the adjacent blobs are equidistant from each other.
Experiment 2: You distribute charge $+6Q$ uniformly on the surface of the sphere. 

\vspace{-1cm}

\begin{enumerate}[Problem 1.]
\setcounter{enumi}{13}
\item \label{problem14} In experiments 1 and 2, points A, B, and C are
equidistant from the center and lie in the same equatorial plane of the
sphere. In experiment 1, points A and C are straight out from two of the
charges and point B is in between points A and C as shown. Which one of the
following statements is true about the electric field magnitudes at points
A, B, and C due to the $+6Q$ surface charge?
\begin{enumerate}[(a)]
\setlength{\itemsep}{-0.1ex plus 0.1ex minus 0.1ex}

\item In each experiment, the field magnitude is the same at points A, B,
and C, but the magnitudes differ in the two experiments.
\item In each
experiment, the field magnitude is the same at points A, B, and C, and the
magnitudes are equal in the two experiments.
\item In experiment 1, the
field magnitude is the same at points A, B, and C, but not in experiment 2.
\item In experiment 2, the field magnitude is the same at points A,
B, and C, but not in experiment 1.
\item None of the above.
\end{enumerate}

\item \label{problem15}
Which one of the following is a true statement about the electric field
magnitude inside the hollow insulating sphere due to the $+6Q$
surface charge (see Fig.~\ref{fig14})?
\begin{enumerate}[(a)]
\setlength{\itemsep}{-0.1ex plus 0.1ex minus 0.1ex}
\item It is zero everywhere inside the sphere in both experiments.
\item It is nonzero everywhere inside the sphere in both experiments.
\item In experiment 1, it has a magnitude that varies from point to point
inside the sphere, but it is zero everywhere inside the sphere in experiment
2.
\item In experiment 1, it has the same nonzero magnitude everywhere
inside the sphere, but it is zero everywhere inside the sphere in experiment
2.
\item None of the above.
\end{enumerate}

\item
\label{prob16} Six positive point charges, $+Q$ each, are placed on an isolated hollow
insulating sphere such that the adjacent point charges are equidistant
(same arrangement as in experiment 1 in Problem~\ref{problem15}). A
spherical and a cubic Gaussian surface concentric with the insulating sphere
are shown in Fig.~\ref{fig16}:
\vspace*{-.2in}
\begin{center}
\begin{figure}[h]
\includegraphics[width=2.5in]{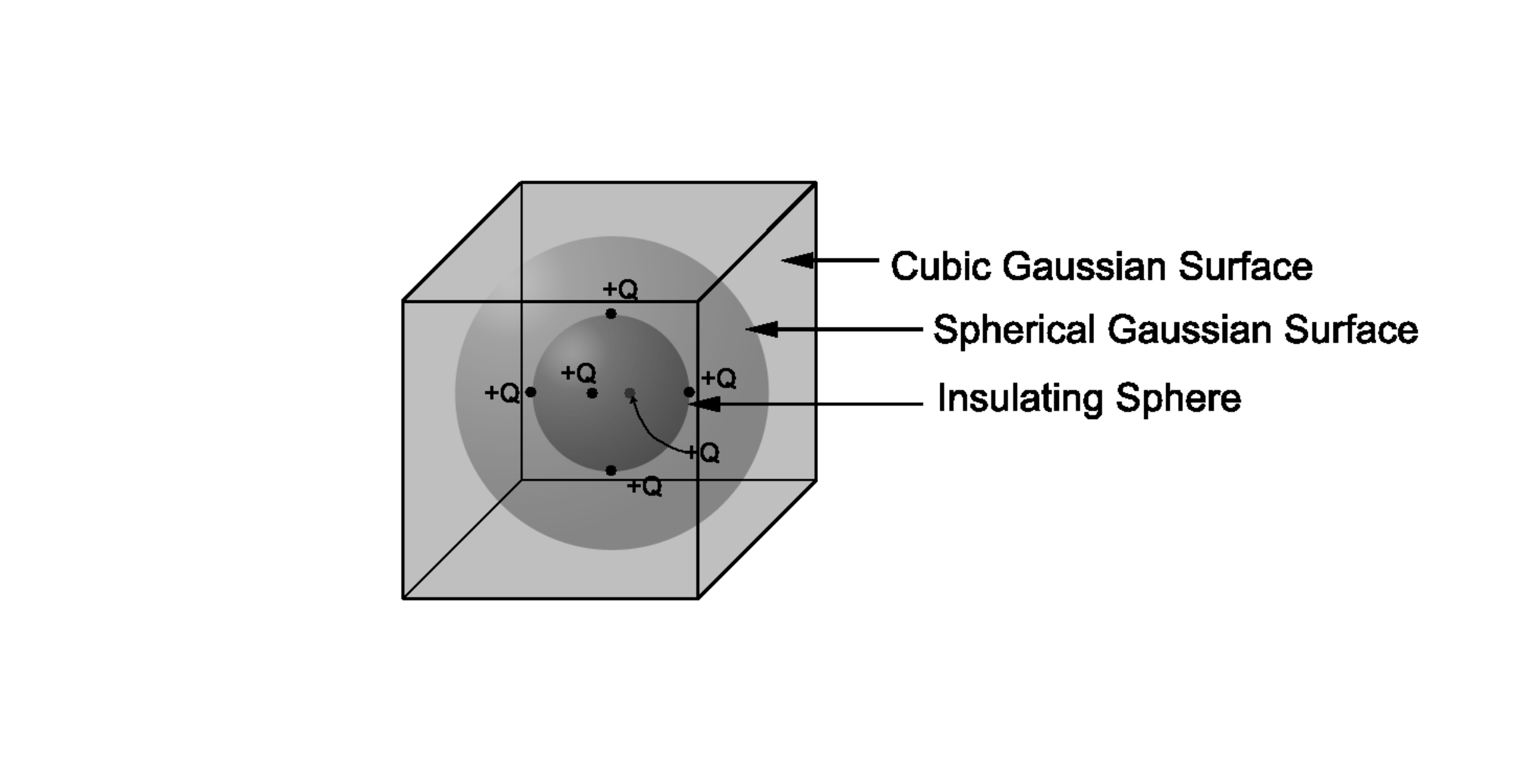}
\caption{\label{fig16}Diagram for Problem~\ref{prob16}.}
\end{figure}
\end{center}
\vspace*{-.4in}
Choose {\em all} of the following statements that are true about the electric
field due to this charge distribution:
\vspace*{-.15in}
\begin{enumerate}[(i)]
\setlength{\itemsep}{-0.1ex plus 0.1ex minus 0.1ex}
\setlength{\itemindent}{1cm}
\item The electric field magnitude is the same everywhere on the
cubic Gaussian surface because the cube has the same symmetry as
that of the charge distribution.
\item The electric field magnitude is the same everywhere on the
spherical Gaussian surface because the sphere has the same
symmetry as the insulating sphere.
\item The electric field is radially outward (straight out from the center) everywhere on the
spherical Gaussian surface. 
\end{enumerate}
\begin{enumerate}[(a)]
\setlength{\itemsep}{-0.1ex plus 0.1ex minus 0.1ex}

\item (i) only.
\item (ii) only.
\item (iii) only.
\item (ii) and (iii) only.
\item None of the above.
\end{enumerate}

\item
\label{prob17} Shown below in Fig.~\ref{fig17} are three thin-walled insulating objects with a net charge $+Q$ uniformly distributed 
on their surfaces: a cube, a sphere, and an open ended cylinder of length $L$ (no caps) and diameter $L$.
The objects are distant from each other so that each may be considered electrically isolated. 
\vspace*{-.11 in}
\begin{center}
\begin{figure}[h]
\includegraphics[width=2.5in]{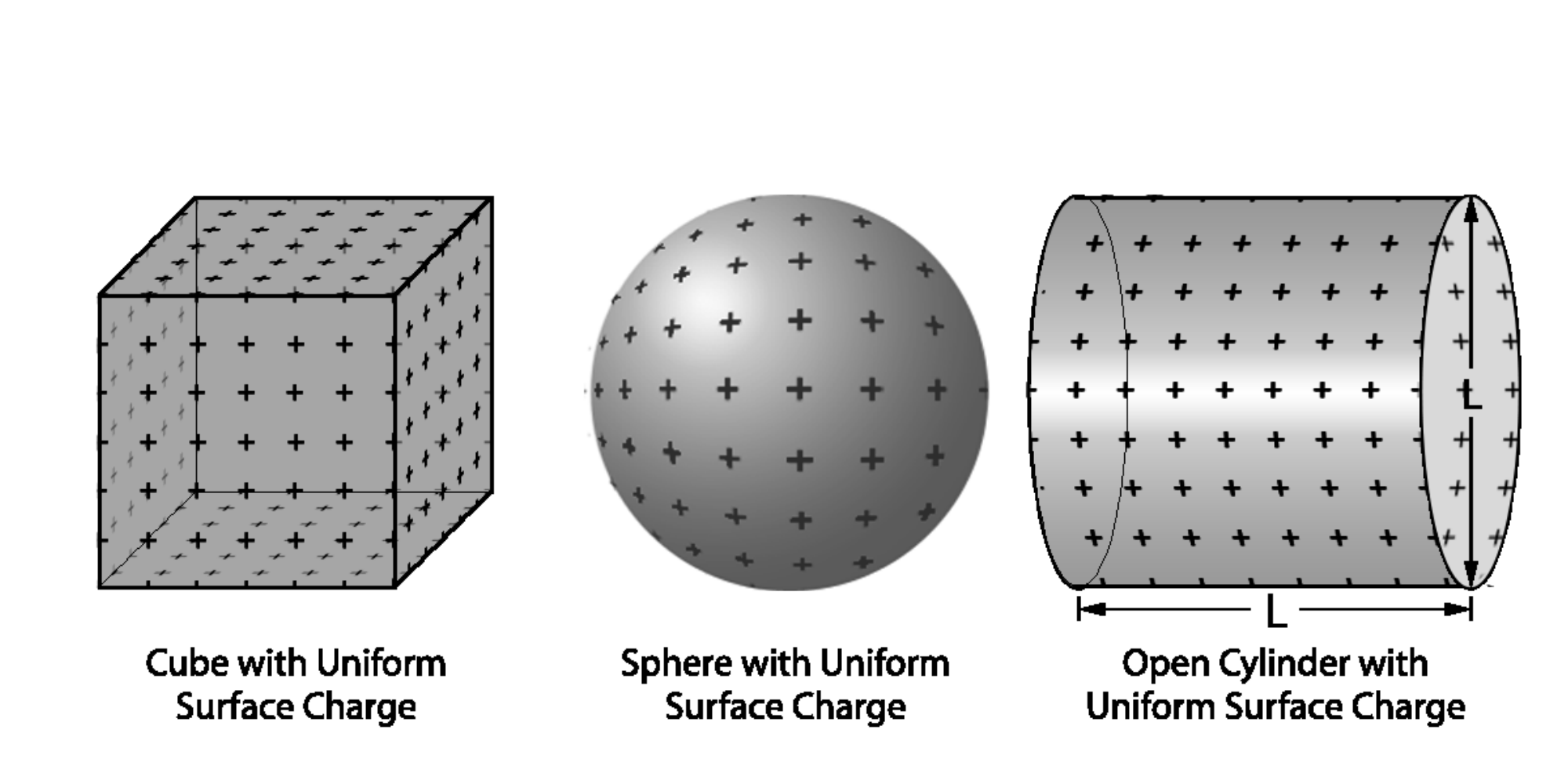}
\caption{\label{fig17} Diagram for Problem~\ref{prob17}.}
\end{figure}
\end{center}
\vspace*{-.22 in}
We can easily use Gauss's law 
to find the electric field due to the uniform surface charge at a point
outside due to:
\begin{enumerate}[(a)]
\setlength{\itemsep}{-0.1ex plus 0.1ex minus 0.1ex}
\item the cube only,
\item the sphere only,
\item the sphere and cylinder only,
\item the sphere and cube only,
\item all three objects,
\end{enumerate}

\item Choose {\em all} of the following statements that are true (Note: This
question does not refer
to a particular charge distribution so a statement is true only if there are
no exceptions):
\vspace*{-.15in}
\begin{enumerate}[(i)]
\setlength{\itemsep}{-0.1ex plus 0.1ex minus 0.1ex}
\setlength{\itemindent}{1cm}
\item If the electric field at every point on a Gaussian surface is zero, the net
electric flux through the surface {\it must} be zero.
\item If there is no charge enclosed inside a Gaussian surface, the electric
field everywhere on the surface {\it must} be zero.
\item If the net electric flux through a Gaussian surface is zero, the
electric field everywhere on the surface {\it must} be zero.
\end{enumerate}
\begin{enumerate}[(a)]
\setlength{\itemsep}{-0.1ex plus 0.1ex minus 0.1ex}
\item (i) only.
\item (ii) only.
\item (i) and (ii) only.
\item (i) and (iii) only.
\item (ii) and (iii) only.
\end{enumerate}

\end{enumerate}

\begin{center}
\begin{figure}[h]
\includegraphics[width=3.3in]{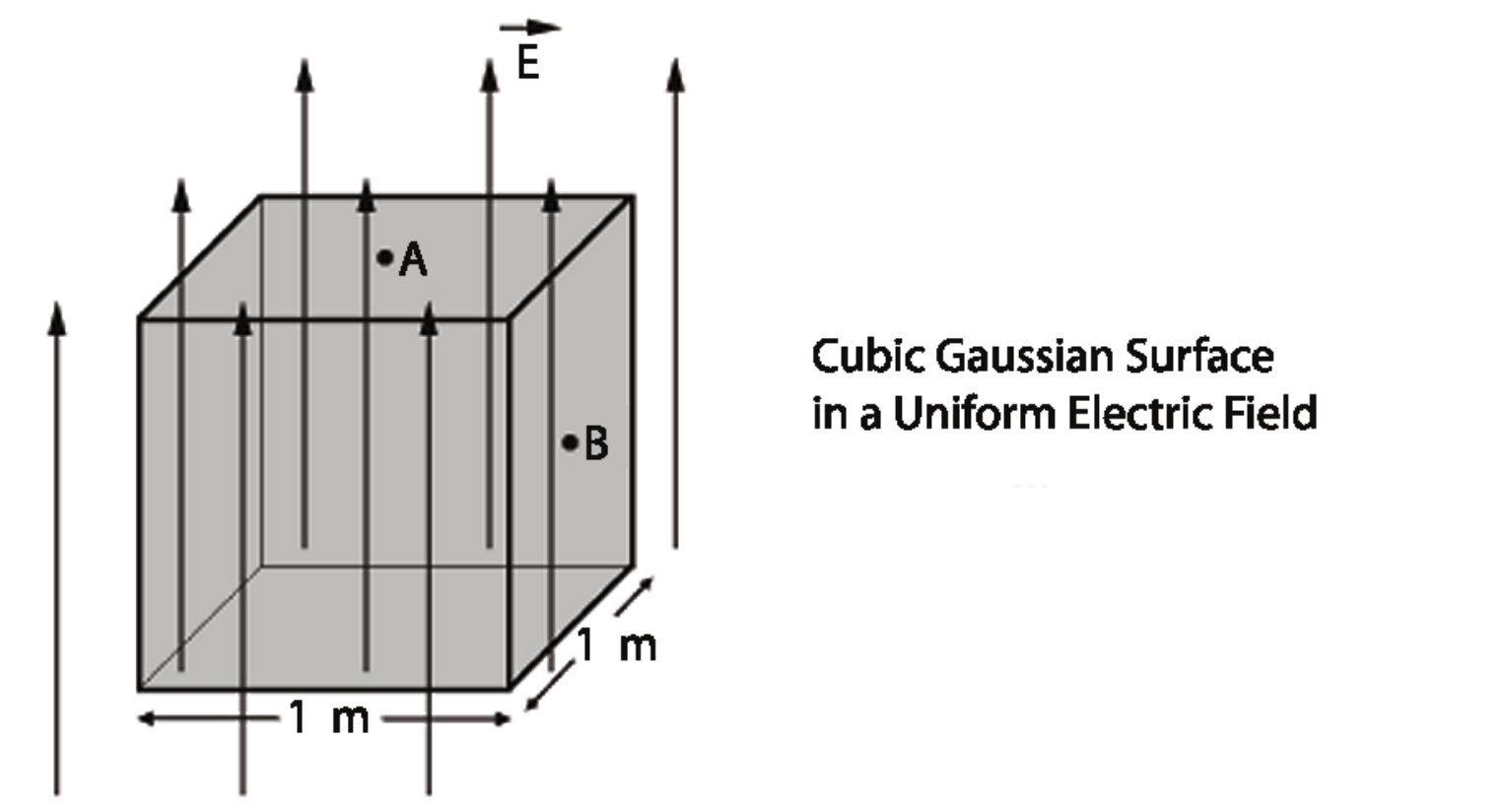}
\caption{\label{fig19} Diagram for Problems~\ref{problem19} and \ref{problem20}.}
\end{figure}
\end{center}

\noindent {\bf Setup for Problems~\ref{problem19} and \ref{problem20}}\\
A cubic Gaussian surface with 1\,m on a side is oriented with two horizontal
and four vertical faces, as shown in Fig.~\ref{fig19}. It is in a uniform electric field of
20\,N/C which is directed vertically upward. Point A is on the
top surface and point B on a side surface of the cubic Gaussian surface.

\vspace*{-.15in}
\vspace*{-.25in}

\begin{enumerate}[Problem 1.]
\setcounter{enumi}{18}

\item \label{problem19}
Which one of the following statements is true about the electric field at
points A and B?
\begin{enumerate}[(a)]
\setlength{\itemsep}{-0.1ex plus 0.1ex minus 0.1ex}
\item The field is zero at both points A and B.
\item The field is zero at point A but not at point B.
\item The field is zero at point B but not at point A.
\item The field is nonzero at both points A and B and its direction is the
same at the two points.
\item The field is nonzero at both points A and B
but its direction is different at the two points.
\end{enumerate}

\item \label{problem20}
Choose all of the following statements that are true about the electric flux.
\vspace*{-.15in}
\begin{enumerate}[(i)]
\setlength{\itemsep}{-0.1ex plus 0.1ex minus 0.1ex}
\setlength{\itemindent}{1cm}
\item The net flux through the whole cubic surface is zero.
\item The magnitude of the flux through the top face of the cubic surface
is 20\,N m$^2$/C.
\item The magnitude of the net flux through the whole cubic surface is
20\,N m$^2$/C.
\end{enumerate}
\begin{enumerate}[(a)]
\setlength{\itemsep}{-0.1ex plus 0.1ex minus 0.1ex}
\item (i) only.
\item (ii) only.
\item (iii) only.
\item (i) and (ii) only.
\item (ii) and (iii) only.
\end{enumerate}

\end{enumerate}

\noindent {\bf Setup for Problems~\ref{problem21} and
\ref{problem22}}

\noindent Shown below in Fig.~\ref{fig21} are four imaginary surfaces coaxial with an isolated infinitely
long line of charge (with uniform linear charge density $\lambda$
$C/m$):
\vspace*{-.15in}
\begin{enumerate}[(i)]
\setlength{\itemsep}{-0.1ex plus 0.1ex minus 0.1ex}
\setlength{\itemindent}{1cm}
\item a closed cylinder of length $L$.
\item a sphere of diameter $L$.
\item a closed cubic box with side $L$.
\item a two dimensional square sheet with side $L$. The plane of the sheet is perpendicular to the line of charge.
\end{enumerate}
\vspace*{-.25in}
\begin{center}
\begin{figure}[h]
\includegraphics[width=1.95in]{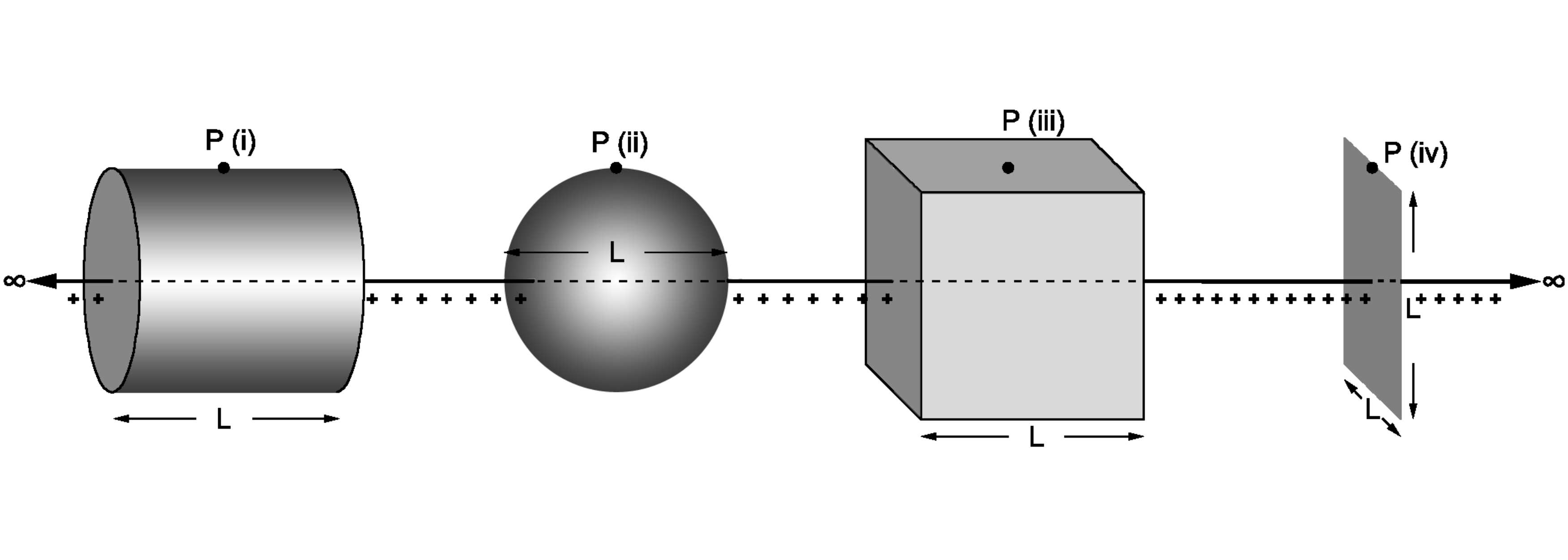}
\caption{\label{fig21} Diagram for Problems~\ref{problem21} and \ref{problem22}.}
\end{figure}
\end{center}
\vspace*{-.3in}

\begin{enumerate}[Problem 1.]
\setcounter{enumi}{20}

\item \label{problem21}
Choose all of the above surfaces through which the net
electric flux is $\Phi_S=\lambda L/\epsilon_0$:
\begin{enumerate}[(a)]
\setlength{\itemsep}{-0.1ex plus 0.1ex minus 0.1ex}
\item (i) only.
\item (i) and (ii) only.
\item (i) and (iii) only.
\item (i), (ii), and (iii) only.
\item (i), (ii), (iii), and (iv).
\end{enumerate}

\item \label{problem22}
Choose all of the above surfaces which can be used as Gaussian surfaces to
easily find the electric field magnitude (due to the infinite
line of charge) at a point P shown on the surface using Gauss's law:
\begin{enumerate}[(a)]
\setlength{\itemsep}{-0.1ex plus 0.1ex minus 0.1ex}
\item (i) only.
\item (i) and (ii) only.
\item (i) and (iii) only.
\item (i), (ii), and (iii) only.
\item (i), (ii), (iii), and (iv).
\end{enumerate}

\item \label{problem23}
The diagram in Fig.~\ref{fig23} below shows the electric field lines in a region.
Sadly, you do not know the field inside the three regions i, ii, and iii.
This cross-sectional drawing is qualitatively correct. 
\vspace*{-.12in}
\begin{center}
\begin{figure}[h]
\includegraphics[width=1.8in]{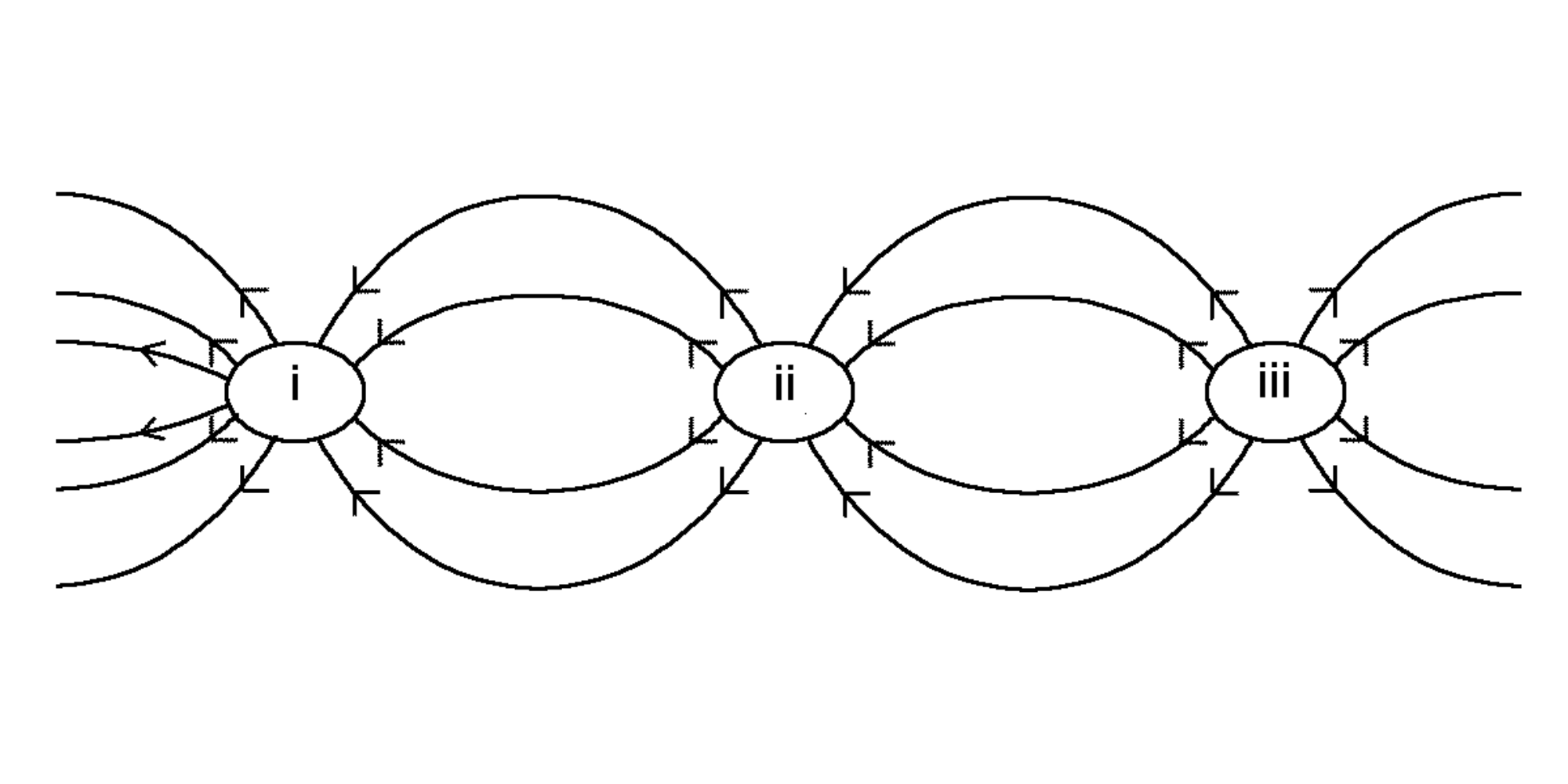}
\caption{\label{fig23} Diagram for Problem~\ref{problem23}.}
\end{figure}
\end{center}
\vspace*{-.21in}
Which region (or regions) carries net charge of the greatest magnitude?
\begin{enumerate}[(a)]
\setlength{\itemsep}{-0.1ex plus 0.1ex minus 0.1ex}
\item (i) only.
\item (ii) only.
\item (iii) only.
\item (ii) and (iii) which have equal net charge.
\item (i), (ii), and (iii) which have equal net charge.
\end{enumerate}

\item 
\label{problem24} Shown below in Fig.~\ref{fig24} are four regions A, B, C, and D (separated by spherical surfaces).
The electric field is zero in regions A (innermost) and D (outermost). The electric field in regions
B and C is radially outward and inward, respectively.
\vspace*{-.12in}
\begin{center}
\begin{figure}[h]
\includegraphics[width=2.55in]{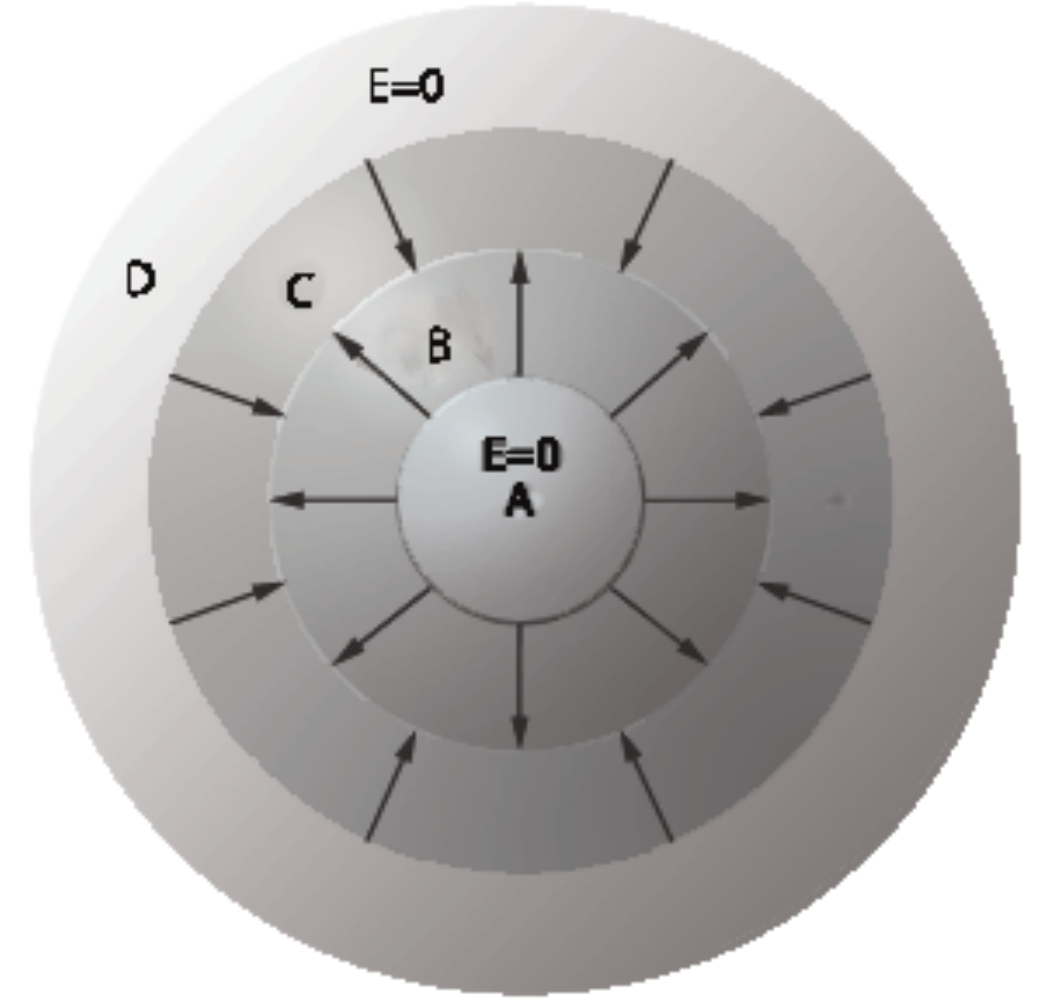}
\caption{\label{fig24} Diagram for Problem~\ref{problem24}.}
\end{figure}
\end{center}
\vspace*{-.05in}
Choose all of the following statements that must be true:
\vspace*{-.15in}
\begin{enumerate}[(i)]
\setlength{\itemsep}{-0.1ex plus 0.1ex minus 0.1ex}
\setlength{\itemindent}{1cm}
\item The combined net charge enclosed in all the regions shown must be zero.
\item There {\it cannot} be a point charge at the center of region A.
\item There must be a negative surface charge between regions B and C.
\end{enumerate}
\begin{enumerate}[(a)]
\setlength{\itemsep}{-0.1ex plus 0.1ex minus 0.1ex}
\item (i) only.
\item (i) and (ii) only.
\item (i) and (iii) only.
\item (ii) and (iii) only.
\item (i), (ii), and (iii).
\end{enumerate}

\item
\label{problem25} In Fig.~\ref{fig25} below, a point charge $+Q$ is near a thin hollow insulating (non-conducting) sphere of radius $L$
that has an equal amount of charge $+Q$ uniformly distributed
on its surface. No other charges are around.
\vspace*{-.14in}
\begin{center}
\begin{figure}[h]
\includegraphics[width=2.37in]{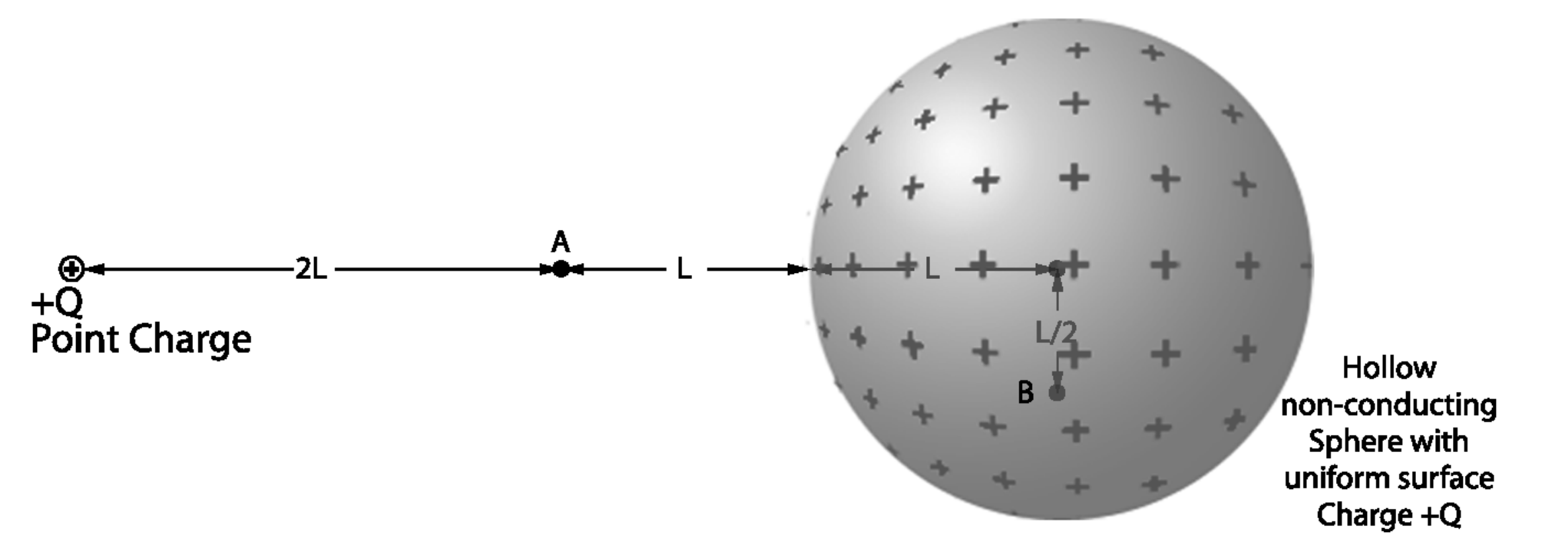}
\caption{\label{fig25} Diagram for Problem~\ref{problem25}.}
\end{figure}
\end{center}
\vspace*{-.05in}
Which one of the following is a true statement about the net electric field (due to the point charge and 
the surface charge on the hollow insulating sphere) at points A (outside the sphere at a distance $2L$ from the center) 
and B (inside the sphere at a distance $L/2$ from the center)?
\begin{enumerate}[(a)]
\setlength{\itemsep}{-0.1ex plus 0.1ex minus 0.1ex}

\item The electric field is zero at point A but is nonzero at point B.
\item The electric field is nonzero at point A but is zero at point B.
\item The electric field is nonzero at both points A and B.
\item The electric field is zero at both points A and B.
\item It is impossible to answer this question without knowing the numerical
value of $Q$.
\end{enumerate}

\end{enumerate}

[xx need last page \# for all articles xx]

\end{document}